\documentclass[12pt]{article}

\usepackage{graphicx}

\begin{document}

%%%%%%%%%%%%%%%%%%%%%%%%%%%%%%%%%%%%%%%%%%%%%%%%%%%%%%%%%%%%%%%%%%%
% sugawara's macro
%%%%%%%%%%%%%%%%%%%%%%%%%%%%%%%%%%%%%%%%%%%%%%%%%%%%%%%%%%%%%%%%%%%
\newcommand{\Om}{\Omega}
\newcommand{\df}{\stackrel{\rm def}{=}}
\newcommand{\co}{{\scriptstyle \circ}}
\newcommand{\de}{\delta}
\newcommand{\lb}{\lbrack}
\newcommand{\rb}{\rbrack}
\newcommand{\rn}[1]{\romannumeral #1}
\newcommand{\msc}[1]{\mbox{\scriptsize #1}}
\newcommand{\dsp}{\displaystyle}
\newcommand{\scs}[1]{{\scriptstyle #1}}

\newcommand{\ket}[1]{| #1 \rangle}
\newcommand{\bra}[1]{| #1 \langle}
\newcommand{\vac}{| \mbox{vac} \rangle }

\newcommand{\e}{\mbox{{\bf e}}}
\newcommand{\va}{\mbox{{\bf a}}}
\newcommand{\bc}{\mbox{{\bf C}}}
\newcommand{\br}{\mbox{{\bf R}}}
\newcommand{\bz}{\mbox{{\bf Z}}}
\newcommand{\bq}{\mbox{{\bf Q}}}
\newcommand{\bn}{\mbox{{\bf N}}}
\newcommand {\eqn}[1]{(\ref{#1})}

\newcommand{\cp}{\mbox{{\bf P}}^1}
\newcommand{\n}{\mbox{{\bf n}}}
\newcommand{\sbz}{\msc{{\bf Z}}}
\newcommand{\sn}{\msc{{\bf n}}}

\newcommand{\be}{\begin{equation}}\newcommand{\ee}{\end{equation}}
\newcommand{\bea}{\begin{eqnarray}} \newcommand{\eea}{\end{eqnarray}}
\newcommand{\ba}[1]{\begin{array}{#1}} \newcommand{\ea}{\end{array}}

\newcommand{\cleqn}{\setcounter{equation}{0}}
\makeatletter
\@addtoreset{equation}{section}
\def\theequation{\thesection.\arabic{equation}}
\makeatother

\def\npb{Nucl. Phys. {\bf B}}
\def\plb{Phys. Lett. {\bf B}}
\def\mpla{Mod. Phys. {\bf A}}
\def\ijmpa{Intern. J. Mod. Phys. {\bf A}}
\def\cmp{Comm. Math. Phys.}
\def\prd{Phys. Rev. {\bf D}}

\def\vu{\vec u}
\def\vs{\vec s}
\def\vv{\vec v}
\def\vt{\vec t}
\def\vn{\vec n}
\def\ve{\vec e}
\def\vp{\vec p}
\def\vk{\vec k}
\def\vx{\vec x}
\def\vz{\vec z}
\def\vy{\vec y}
\def\vxi{\vec\xi}

%\newcommand{\matriz}[4]{\left(\begin{array}{cc}#1&#2\\#3&#4\end{array}\right)}

%%%%%%%%%%%%%%%%%%%%%%%%%%%%%%%%%%%%%%%%%%%%%%%%%%%%%%%%%%%%%%%%%%%
\begin{flushright}
La Plata Th-06/01\\December 2006\\Revised version, April 2007
\end{flushright}

\bigskip

\begin{center}

{\Large\bf Space-time filling branes in non critical (super) string
theories}
\footnote{
This work was partially supported by CONICET,Argentina. }
\bigskip
\bigskip

{\it \large Adri\'{a}n R. Lugo and Mauricio B. Sturla}
\footnote{ {\sf
lugo@fisica.unlp.edu.ar, $\;\;$ sturla@fisica.unlp.edu.ar} }
\bigskip

{\it Departamento de F\'\i sica, Facultad de Ciencias Exactas\\
Universidad Nacional de La Plata\\ C.C. 67, (1900) La Plata, Argentina
}
\bigskip
\bigskip

\end{center}

\begin{abstract}
We consider solutions of (super) gravities associated to non-critical
(super) string theories in arbitrary space-time dimension
$D=p+3$, that describe generically non extremal black $p$-branes
charged under NSNS or RR gauge fields, embedded in some non
critical vacuum. In the case of vacuum (uncharged) backgrounds, we
solve completely the problem obtaining {\it all} the possible
solutions, that consist of the $(p+1)$-dimensional Minkowski space
times a linear dilaton times a $S^1$, and a three parameter family
of solutions that include $(p+1)$-dimensional Minkowski space
times the cigar, and its T-dual $(p+1)$-dimensional Minkowski
space times the trumpet. For NSNS charged solutions, we also solve
in closed form the problem, obtaining several families of
solutions, that include in particular the fundamental non-critical
string solution embedded in the cigar vacuum, recently found in
hep-th/0604202, a solution that we interpret as a fundamental non-critical
string embedded in the linear dilaton vacuum, and a two-parameter family of regular
curvature solutions asymptotic to $AdS_{1,2}\times S^1$.
In the case of RR charged $Dp$-branes solutions, an ans\"atz allows us to find a non conformal,
constant curvature, asymptotically $AdS_{1,p+1}$ space, T-dual to $AdS_{1,p+2}$, together with a
two-parameter family of solutions that includes the non conformal, $AdS$ black hole like solution
associated with the earlier space.
The solutions obtained by T-duality are Einstein spaces consisting of a two-parameter family of
conformal, constant dilaton solutions, that include, in particular, the AdS black hole of hep-th/0403254.
We speculate about the possible applications of some of them in the framework
of the gauge-gravity correspondence.

\end{abstract}

\section{Introduction}
\cleqn

The AdS/CFT correspondence and further generalizations
\cite{Aharony:1999ti} have turned to be an interesting way of
addressing the old problem of solving the low energy dynamics of
gauge theories, through the realization of the holographic ideas in
the context of superstring theories in non trivial vacua. So far,
the most interesting results were obtained mainly for supersymmetric
(SUSY) field theories, in the supergravity (SUGRA) limit
\cite{Edelstein:2006kw}. In reference \cite{Witten:1998zw}, a
possible mechanism to get holographic gravity duals of non
supersymmetric gauge theories was proposed, by compactifying
supersymmetric systems in higher dimensions with antiperiodic, SUSY
breaking, boundary conditions for the fermions. However, the
holographic set-up in all these cases suffers from several
limitations, derived from the impossibility, at the present, to
quantize strings on Ramond-Ramond (RR) backgrounds. Due to this
fact, a low energy, SUGRA approximation becomes necessary. On the
other hand, the existence of transverse, compact spaces, leads to
Kaluza-Klein (KK) towers of states with masses of order of the gauge
theory scale (and then, of the scale of hadronic states), that
therefore do not decouple from the theory and contaminate the
spectrum. The introduction of non critical backgrounds (in
dimensions $D<10$) provides a natural way to overcome this problem.
The idea to extend the holographic description of gauge theories to
non critical backgrounds was first introduced by Alexander Polyakov
\cite{Polyakov:1998ju}, who proposed a dual of pure Yang-Mills (YM)
theories in terms of a five dimensional non critical gravity
background.

Non critical string theories in $D$ dimensional space-time are
characterized by including on their world-sheet (at least) one scalar field,
the Liouville mode, which combines together with additional
$D-1$ coordinates \cite{Polyakov:1981rd}, \cite{Polyakov:1981re}.
Much work was made on the subject in the past, in particular in $D\leq 2$ and
the related matrix models \cite{Ginsparg:1993is}.
However, the so called ``$c=1$ barrier"  seemed to forbid to go to higher
dimensions \cite{Seiberg:1990eb}.
Things change dramatically with the introduction of $({\cal N}_L,{\cal N}_R)=(2,2)$
superconformal symmetry on the world-sheet.
In reference \cite{Kutasov:1990ua}, it
was showed that non critical type II superstring theories can be formulated
in $d=2n , n=0,1,\dots,4$ space-time dimensions, and describe
consistent solutions of string theory in sub-critical dimensions,
with space-time supersymmetry consisting of, at least, $2^{n+1}$ supercharges.
On the world-sheet, these theories present, in addition to the dynamical Liouville mode
$\phi$ mentioned above, a compact boson $X$, being the target space of the general form
\be
{\cal M}^d\times \Re_\phi\times S^1 \times M/\Gamma\qquad ,
\ee
where ${\cal M}^d$ is $d$-dimensional Minkowski space-time, $M$ is
an arbitrary $(2,2)$ superconformal field theory, and $\Gamma$ a
discrete subgroup acting on $S^1 \times M$.
Now, the Liouville theory include a linear dilaton background that yields a strong
coupling singularity; however the existence of the compact boson
allow to resolve this singularity by replacing the $\Re_\phi\times
S^1$ part of the background by the ${\cal N}=2$ Kazama-Suzuki
supercoset $SL(2,\Re)_k/U(1)$.
This space has a two-dimensional cigar-shaped geometry, with a natural scale given by
$\sqrt{k\,\alpha'}$ \cite{Witten:1991yr}, \cite{Mandal:1991tz}.
It provides a geometric cut-off for the strong coupling singularity,
while coinciding with the linear dilaton solution in the weak coupling region.
It is known that this vacuum solution is an exact solution to all
orders in type II theories, up to a trivial shift $k\rightarrow k-2$
\cite{Bars:1992sr}.
When $d=8$, we go back to the critical superstring in flat ten dimensional
Minkowski space-time.

In view of this, we will consider in this paper backgrounds of type II non critical
string theories that include a two dimensional non trivial sector parameterized by
a radial coordinate and a $S^1$ coordinate, the radial coordinate having the interpretation
of a energy scale from the field theory point of view \cite{Polyakov:1998ju},
\cite{Maldacena:1997re}, \cite{Itzhaki:1998dd}.
In reference \cite{Lugo:2005yf} the solution of a fundamental non critical string placed
at the tip of the cigar and localized also at the origin of the transverse space was presented.
We will consider here general $p$-branes with different charges that fills all
Minkowski space (no transverse space).

Some papers addressing similar problems have appeared in the
literature, most notably reference \cite{Kuperstein:2004yk} (see
also \cite{Klebanov:2004ya}, \cite{Alishahiha:2004yv}, \cite{Bigazzi:2006ix}).
But, rather than approach the problem through a set of BPS equations
(hopefully supersymmetric) derived from a superpotential that
relates to the potential of the system, we prefer to solve
systematically the full system of second order equations of motion.

An important point to remark is the following one. The fact of
being outside criticality leads, in some cases, to solutions
trustable in a large region of the space, as it is the case of
\cite{Lugo:2005yf} and some solutions presented in this paper,
much in the same way as it happens with the $p$-brane solutions of
critical theories. In other cases, like in $AdS$ type solutions,
the curvature in string units is of order unity, and it is not
clear if they receive important corrections from the higher order
terms in the effective action. We adopt, along the lines followed
in  \cite{Polyakov:1998ju}, \cite{Klebanov:2004ya}, \cite{Kuperstein:2004yk},
\cite{Kuperstein:2004yf}, \cite{Casero:2005se}, the posture that
for special solutions like the  $AdS$ like ones, the conformal structure of
the background would not be modified  by the higher order
curvature contributions but only the corresponding parameters will get
renormalized.

The paper is organized as follows.
In Section $2$ we present the non-critical low energy effective action, and the equations of motion corresponding
to the space-time filling $p$-brane ans\"atz to be considered, and reduce the full system of second order differential
equations to a pair of coupled equations plus a constraint, ``zero energy" condition, equations
(\ref{system}) and (\ref{constraint}) respectively.
In Section $3$ we obtain all the possible vacuum solutions.
They consist of the solution given by Minkowski space-time times a linear dilaton times $S^1$,
plus a three-parameter family asymptotic to it but singular in general, excluding a two-parameter sub-family
(that includes the well-known Minkowski space-time times the cigar), regular in the sense that it presents bounded
both the string coupling constant and the scalar curvature.
In Section $4$ we solve completely the problem for NSNS charged solutions.
Other than the well-known $AdS_{1,2}\times S^1$ space-time (represented by the exact $SL(2,\Re)\times U(1)$ model),
there are several families of solutions which have it as an asymptotic limit.
Most notably, a solution interpretable as a fundamental non critical string embedded in the linear dilaton vacuum,
and a two-parameter family of regular solutions.
Moreover we recover the solution of fundamental string in the cigar vacuum recently found,
and we also get three families of oscillating, singular solutions, with no obvious interpretation.
In Section $5$ we consider the system for RR charged solutions; under an assumption
we are able to get a two-parameter family of solutions asymptotic to a space-time T-dual of $AdS_{1,p+2}$ space,
singular in the infrared limit except two regular solutions.
Nevertheless, via a T-duality transformation, it maps into a conformal, constant dilaton family of Einstein spaces
that includes the $AdS$ black hole of \cite{Kuperstein:2004yk}.
In Section $6$ we draw the conclusions and future perspectives.
An appendix that collects the formulae used in the computations is added at the end.

\section{The general setting.}
\cleqn

\subsection{The non critical action and the space-time filling $p$-brane ans\"atz.}

Our starting point is the bosonic part of the low energy effective action of non critical
(super) strings in $D$ dimensions, that in string frame reads,
\begin{eqnarray}
S[\Psi]&=& \frac{1}{2\,\kappa_D{}^2}\;\int \epsilon_G \;e^{-2\Phi}\left(
R[G]+4\left(D\Phi\right)^2+\Lambda^2-{1\over 2}\sum_q
e^{\left({b_q}+2\right)\Phi}\left(F_{q+2}\right)^2 \right)\qquad.\label{sugraction}
\end{eqnarray}
Here $\Psi$ stands for the fields $\{G_{mn},\Phi, A_{q+1}\}$, $F_{q+2} = dA_{q+1}$ is the
field strength of the gauge field form $A_{q+1}$, and
$\epsilon_G= \omega^0\wedge\dots\wedge\omega^{D-1}=d^D x\,\sqrt{-\det G}\, $
is the volume element.
The constant $b_q$ is equal to $-2 (0)$ for NSNS (RR) forms,
$\Lambda^2$ is a cosmological constant (assumed positive) that we identify in (super) string theories with
($\frac{10-D}{\alpha'}$) $\frac{2\,(26-D)}{3\,\alpha'}$, and $\kappa_D{}^2$ is the
$D$-dimensional Newton constant.
Furthermore, we assume a source term of the form,
\be
S^{source}[\Psi]= \sum_q\; \mu_q\, \int A_{q+1}\wedge* J_{q+1}\label{source}
\ee
where $\mu_q$ is the charge of the source under $A_{q+1}$.
The equations of motion that follow are,
\begin{eqnarray}
R_{mn}&=&-2\,D_mD_n\Phi + T^{A}_{mn}\cr
\Lambda^2 &=& e^{2\Phi}D^2(e^{-2\Phi})+
\sum_q {D\,(2+b_q)-2\,b_q-4(q+2)\over 8}\;e^{(2+b_q)\Phi}\; (F_{q+2})^2\cr
d\left(e^{b_q\,\Phi}*F_{q+2}\right)&=& (-)^q \; Q_q\; *J_{q+1}\;\;\;\;,\;\;\;\;
Q_q\equiv 2\,\kappa_D{}^2\, \mu_q \label{ecformal}
\end{eqnarray}
where the gauge energy-momentum tensor and strength field contractions are
\footnote{
In string theories,  $\kappa_D{}^2 \sim \,\alpha'^{\frac{D-2}{2}}$; the charge of
$N_q$ $q$-branes is $|\mu_q| \sim N_q\,\alpha'^{-\frac{q+1}{2}}$, and so
$|Q_q|\sim N_q\, \alpha'^{\frac{D-q-3}{2}}$.
We do not intend to fix the normalizations in the noncritical context of the present paper.
},
\bea
T^{A}_{mn}&\equiv&\sum_q {1\over2}\, e^{(2+b_q )\Phi}\left( \left(F_{q+2}\right)^2_{mn}
-\frac{2+b_q}{4}\; G_{mn} \;\left(F_{q+2}\right)^2\right)\cr
(F_{q+2})^2_{mn} &\equiv& \frac{1}{(q+1)!}\, G^{m_1 n_1}\dots G^{m_{q+1}n_{q+1}}\;
{F_{q+2}}_{mm_1\dots m_{q+1}}{F_{q+2}}_{nn_1\dots n_{q+1}}\cr
(F_{q+2})^2&\equiv& \frac{1}{(q+2)!}G^{m_1 n_1}\dots G^{m_{q+2}n_{q+2}}\;
{F_{q+2}}_{m_1\dots m_{q+2}}{F_{q+2}}_{n_1\dots n_{q+2}}\cr
%=\frac{1}{q+2}\,G^{mn}\;(F_{q+2})^2_{mn}\cr
& &\label{emt}
\eea

Let us now consider the following ans\"atz for the fields,
\bea
G &=& -A^2\; dx^0{}^2+\tilde{A}^2\;d{\vec x}{}^2 + C^2\;d\rho^2 +\tilde C^2\; dz^2\cr
A_{p+1} &=& dx^0\wedge\dots\wedge dx^p\; E(\rho)\cr
\Phi &=& \Phi(\rho)
\label{fieldansatz}
\eea
where we have assumed that there is one type of charge, corresponding
to the $A_{p+1}$ form (only $q=p$ is present).
This ans\"atz would correspond to a black $p$-brane extended along $( x^0, \vec x)$,
and localized along the radial variable $\rho$ in the transverse two dimensional
euclidean space with coordinates ($\rho, z$), preserving $SO(2)$ rotational
invariance in the $S^1$ variable $z$, whose compactification radius $R$, $z \sim z + 2\,\pi\, R $, is determined
in most cases by the charges.
The vacuum solution in which it is embedded, in view of the non zero cosmological constant
$\Lambda$, must be non trivial (see Section $3$).
In the appendix we collect relevant formulae related with this ans\"atz.

\subsection{The equations of motion and the general solution.}

When a p-brane fills all the Minkowski space, the transverse space is just
two-dimensional and the solution, admitting a $SO(2)$ invariance, depends on the ``radial"
coordinate $\rho$ of the deformed space (linear dilaton, cigar, etc, see Section $3$).
By realizing that $C(\rho)$ is defined up to a $\rho$-reparameterization
(in fact, it transforms as a one form, all the other metric functions being scalar),
it is natural to search for a scalar variable under such diffeomorphisms.
This is accomplished by the introduction of the coordinate $x$, defined in the following way
\footnote{
Unless stated different, we will be considering $x$ non negative along this paper.
},
\bea
x &=& \int^{\rho}\; \frac{d\rho}{H(\rho)} \;\;\;\;,\;\;\;\;\partial_x=H(\rho)\;\partial_{\rho}\cr
H&\equiv& \frac{F_1\; e^{-2\Phi} }{C^2} = A\;\tilde A^p\; \tilde C\;e^{-2\Phi}\;C^{-1}
\label{xdef}
\eea
Let us notice that  $H$ is a vector and $F_1$ another $1$-form, and $H\,C$ a scalar.
Furthermore,
\be
C(\rho)\; d\rho = H\;C\;dx
\ee
an useful relation to write down the metric.
In terms of this coordinate, and denoting with a prime a $x$-derivative,
the equations of motion (\ref{1})-(\ref{6}) take the following form,
\bea
(\ln A)''&=& \frac{2-b_p}{8}\;{e^{(2+b_p)\Phi}\over (A{\tilde A}^p)^2}\; E'^2\label{A}\\
(\ln \tilde A)''&=& \frac{2-b_p}{8}\;{e^{(2+b_p)\Phi}\over (A{\tilde A}^p)^2}\; E'^2
\label{tildeA}\\
(\ln \tilde C)''&=& -\frac{2+b_p}{8}\;{e^{(2+b_p)\Phi}\over (A{\tilde A}^p)^2}\; E'^2
\label{tildeC}\\
(\ln e^{-2\Phi})''&=& \Lambda^2\; (H\;C)^2 - \frac{2-b_p}{8}\,(p+1)\;{e^{(2+b_p)\Phi}\over (
A{\tilde A}^p)^2}\; E'^2\label{Phi}\\
\left(\frac{e^{(2+b_p)\Phi}}{(A{\tilde A}^p)^2}\; E'\right)'&=&
(-)^{p+1} \,Q_p\;\frac{(H\;C)^2}{A{\tilde A}^p}\; e^{2\Phi}\;\delta^2_{G^\perp}
\label{E}\\
-(\ln H\,C)'' + (\ln H\,C)'^2 &=& (\ln A)'^2 + p\, (\ln \tilde A)'^2 + (\ln \tilde C)'^2 -
\frac{2-b_p}{8}\;{e^{(2+b_p)\Phi}\over (A{\tilde A}^p)^2}\; E'^2\label{C}\cr
& &
\label{equations}
\eea
We have written at the end the $C$-equation, that we will take as a constraint equation
coming from the gauge fixing of the coordinate $\rho$, implicitly made through the
introduction of the coordinate $x$.

Let us start to solve these equations. Outside the location of the
source, equation (\ref{E}) is solved by,
\be
E'= q\; \left(A\;\tilde A^p\right)^2\, e^{-(2+b_p) \Phi}
\ee
where $q$ is related to $Q_p$ by the relation (obtained by integration of (\ref{ecformal})),
\be
q = \frac{Q_p}{V_z}\qquad,\qquad V_z =\int\,dz\label{qQ}
\ee
The solution of (\ref{A}) is,
\be
A=e^{\alpha\,x}\; \tilde A
\ee
while that the solution of
(\ref{tildeC}) is,
\be
\tilde C = e^{\gamma\,x}\;\tilde A^{-\frac{2+b_p}{2-b_p}}
\ee
where $\alpha, \gamma,$ are arbitrary constants.
We remain with two equations, (\ref{tildeA}) and (\ref{Phi}), plus the constraint (\ref{C}).
Let us introduce the following functions
\footnote{
We are discarding constants that can be absorbed by trivial redefinitions or re-scalings of the coordinates $(x^\mu , z)$.
%for this reason we have introduced the $V_z$ factor in the definition of $f_2$.
},
\bea
f_1(x)&\equiv& (H C)^2 = \left(A\;\tilde
A^p\; \tilde C\;e^{-2\Phi}\right)^2 = \left(
e^{(\alpha+\gamma)\,x}\; \tilde
A^{p+1-\frac{2+b_p}{2-b_p}}\;e^{-2\Phi}\right)^2\cr
V_z{}^2\; f_2(x)&\equiv&\left({H\,C \over \tilde C}\right)^2\;e^{(2-b_p)\Phi}
= \left(A\;\tilde A^p\right)^2\;e^{-(2+b_p)\Phi}=
\left(e^{\alpha\,x}\;\tilde A^{p+1}\right)^2\;e^{-(2+b_p)\Phi}\cr
& &\label{f12def}
\eea
In terms of them, the solution for the fields is given by,
\bea G &=&
\tilde A^2\;\left(-e^{2\alpha\,x}\; dx^0{}^2 + d\vec x ^2\right) + f_1(x)\;dx^2 + e^{2\gamma\,x}\;\tilde
A^{-2\frac{2+b_p}{2-b_p}}\; dz^2\cr
e^{2\sigma\Phi} &=&
e^{2\left(\frac{2+b_p}{2-b_p}\alpha+(p+1)\gamma\right)\,x}\;
\frac{ (V_z{}^2\;f_2)^{p+1 -\frac{2+b_p}{2-b_p}} }{f_1{}^{p+1}}\cr
F_{p+2}&=& V_z\; Q_p\;f_2(x)\; dx\wedge dx^0\wedge\dots\wedge dx^p\label{gralsolution}
\eea
where
\bea \tilde A^{2\sigma} &=&
V_z{}^4\;e^{
(-(2-b_p)\alpha+(2+b_p)\gamma)\,x}\;\frac{f_2{}^2}{f_1{}^{\frac{2+b_p}{2}}}\cr
\sigma
&\equiv&\frac{1}{2}\left((2-b_p)(p+1)+\frac{(2+b_p)^2}{2-b_p}\right)
\eea
It is not difficult to see that the equations (\ref{tildeA})
and (\ref{Phi}) can be recast in terms of $f_1, f_2$ in the
following form,
\bea
(\ln f_1)''&=&2\,\Lambda^2\; f_1 -{2 +b_p\over
4}\, Q_p{}^2\; f_2\cr (\ln f_2)''&=& \frac{2+b_p}{2}\,\Lambda^2\; f_1 +
\frac{(2-b_p)^2}{16}\,(p+1)\,Q_p^2\;f_2 \label{system}
\eea
while that the constraint (\ref{C}) becomes,
\bea
-\frac{1}{2}\,(\ln
f_1)''+\frac{1}{4}\,(\ln f_1)^{'\:2} &=& \left((\ln\tilde A)' +
\alpha\right)^2 +p\,(\ln\tilde A)'{}^2
+\left(-\frac{2+b_p}{2-b_p}\,(\ln\tilde A)' + \gamma\right)^2\cr
&-& \frac{2-b_p}{8}\, Q_p{}^2\; f_2\cr
2\,\sigma\,(\ln\tilde A)' &=& -
(2-b_p)\,\alpha + (2+b_p)\,\gamma + \left(
\ln\frac{f_2{}^2}{f_1{}^{\frac{2+b_p}{2}}}\right)'\qquad.\label{constraint}
\eea
In conclusion, the general solution (\ref{gralsolution}) for
space-time filling, non critical $p$-branes, is determined by the system
(\ref{system}) and the constraint (\ref{constraint}).

Let us start a systematic survey of the possible solutions with the simplest case.

\section{Uncharged solutions $Q_p=0$; the vacua.}
\cleqn

The uncharged solutions represent the possible noncritical vacuum backgrounds.
The system (\ref{system}) reduces to,
\bea
(\ln f_1)''&=&2\,\Lambda^2\; f_1 \cr
(\ln f_2)''&=& \frac{2+b_p}{2}\,\Lambda^2\; f_1
\eea
Plugging the first equation into the second one we obtain $f_2$ in terms of $f_1$,
\be
f_2(x) = e^{\epsilon_0 +\epsilon\, x}\; f_1(x)^{\frac{2+b_p}{4}}
\ee
where $\epsilon_0, \epsilon$ are arbitrary constants.
To solve for $f_1$, we rewrite the first equation in the form,
\be
(\ln f_1)''=2\,\Lambda^2\; f_1\;\;\longleftrightarrow\;\;
{1\over f_1}{d\over d x}\left({1\over f_1}{d\over d x}f_1(x)\right)=2 \,\Lambda ^2
\;\;\longleftrightarrow\;\; \frac{d^2 f_1}{dy^2}=2 \,\Lambda ^2\label{uncharf1eq}
\ee
where we have introduced the primitive $y(x)$ (relevant up to a constant),
\be
y(x)=\int^x \;dx\; f_1(x)\;\;\longleftrightarrow\;\; f_1(x) = y'(x)
\ee
By trivial integration in (\ref{uncharf1eq}) we get,
\be
f_1(x)= y'(x) = \Lambda^{\:2}\; y(x)^{\:2}+ 2\, \beta \; y(x)+\gamma\label{f1yprime}
\ee
where $\beta, \gamma$ are arbitrary integration constants, whose general solution is,
\be
x - x_0= \int^{y(x)}{dy\over \Lambda^{\:2}\; y^{\:2}+ 2 \,\beta\; y+\gamma}
= \left\{\begin{array}{lcl}
{1\over 2\,\sqrt{-\Delta}}\,\ln\left|{\Lambda^2 y+\beta-\sqrt{-\Delta}\over
\Lambda^2 y+\beta+\sqrt{-\Delta}}\right| \;\;\; &,&\;\;\;\Delta < 0\cr
-{1\over \Lambda^2 y+ \beta}\;\;\; &,&\;\;\;\Delta=0\cr
{1\over \sqrt{\Delta}}\, \arctan{\Lambda ^2 y+ \beta \over \sqrt{\Delta}},
\;\;\; &,&\;\;\;\Delta> 0
\end{array}\right.
\ee
where $\Delta=\Lambda^2\,\gamma - \beta^2$.
Thus, we see there are three possible branches, depending of the sign of $\Delta$.
The corresponding value of $y(x)$ is
\be
\Lambda^2\,y(x) +\beta =
\left\{\begin{array}{lcl}
\begin{array}{lcl}
-\sqrt{-\Delta}\:\coth\left(\sqrt{-\Delta}\,(x-x_0)\right)
\;\;&,&\;\; |\Lambda^2\,y(x) +\beta|>\sqrt{-\Delta}\cr
-\sqrt{-\Delta}\:\tanh\left(\sqrt{-\Delta}\,(x-x_0)\right)
\;\;&,&\;\; |\Lambda^2\,y(x) +\beta|<\sqrt{-\Delta}\cr
\end{array}&,&\Delta < 0\cr
-\frac{1}{x-x_0}&,&\Delta=0\cr
\sqrt{\Delta}\, \tan\left(\sqrt{\Delta}\,(x-x_0)\right)&,&\Delta> 0
\end{array}\right.
\ee
which yields, according to (\ref{f1yprime}),
\be
f_1(x) =
\left\{\begin{array}{lcl}
\begin{array}{lcl}
-{\Delta \over \Lambda^2} \:{1\over \sinh^2\left(\sqrt{-\Delta}\,(x-x_0)\right)}
\;\;&,&\;\; |\Lambda^2\,y(x) +\beta|>\sqrt{-\Delta}\cr
+{\Delta \over \Lambda^2} \:{1\over \cosh^2\left(\sqrt{-\Delta}\,(x-x_0)\right)}
\;\;&,&\;\; |\Lambda^2\,y(x) +\beta|<\sqrt{-\Delta}\cr
\end{array}\;\;&,&\;\;\Delta < 0\cr
\frac{1}{\Lambda^2\,(x-x_0)^2}\;\;&,&\;\;\Delta=0\cr
{\Delta\over \Lambda^2}\:{1\over \cos^2\left(\sqrt{\Delta}\,(x-x_0)\right)}
\;\;&,&\;\;\Delta> 0\qquad .
\end{array}\right.\label{f1sn}
\ee
We see that $f_1$ depend on two free parameters, $x_0$ and $\Delta$, while that the whole
solution (\ref{gralsolution}) depends also non trivially (see below) on
$\alpha, \epsilon, \gamma $.
Let us pass to analyze each branch separately.

\subsection{Solution with $\Delta=0$; the linear dilaton.}

In this case, the constraint equation (\ref{constraint}) enforces the condition
$\alpha=\epsilon=\gamma=0$; from (\ref{f1sn}) we have,
\bea
f_1&=&{1\over \Lambda^2\, (x-x_0)^2}\cr
f_2&=&  e^{\epsilon_0}\;\left({1\over \Lambda^2\, (x-x_0)^2}\right)^\frac{2+b_p}{4}
\eea
By introducing the variable $Y = -\Lambda^{-1}\,\ln(\Lambda\,|x-x_0|)$,
and after trivial re-scalings and redefinitions we get,
\bea
G&=& \eta_{1,p}+d Y^2+ r'_0{}^2\; d\theta^2\cr
\Phi&=&\Phi_0 -\frac{1}{r_0}\,Y\qquad,\qquad r_0 \equiv \frac{2}{\Lambda}
\eea
the direct product of $p+1$ dimensional Minkowski space-time and the
linear dilaton solution times a $S^1$ (assuming $\theta\sim\theta+2\,\pi$) of arbitrary radius $r'_0$.
The linear dilaton is widely discussed in the literature, see \cite{Myers:1987fv}, \cite{polcho1}.
Furthermore, a Wick rotated version of this solution is considered as a cosmological model in \cite{Antoniadis:1990uu}.
\footnote{
For cosmological models in string theory, see also \cite{Tseytlin:1991xk}, \cite{Bergshoeff:2005bt}.
}

\subsection{Solutions with $\Delta<0$; the cigar and more.}

From the definition (\ref{f12def}), $f_1= (H \,C)^2 >0$, and therefore the physically
relevant solution in (\ref{f1sn}) is the first one; then
\bea
f_1 (x) &=& {-\Delta \over \Lambda^2} \:{1\over \sinh^2\left(\sqrt{-\Delta}\,(x-x_0)\right)}\cr
f_2(x)&=& e^{\epsilon_0 +\epsilon\, x}\;
\left({-\Delta \over \Lambda^2} \:{1\over \sinh^2\left(\sqrt{-\Delta}\,(x-x_0)\right)}
\right)^{\frac{2+b_p}{4}}
\eea
After various redefinitions
%By shifting $x$ by $x_0$ (in this case, $x_0$ can be trivially put to zero, but it is not always so, see next subsection),
%re-scaling it by $(-\Delta)^{-\frac{1}{2}}\,$, and redefining the integration constants $\;\alpha;,\;\epsilon\;,\;\gamma\;$,
%conveniently to $\;a\;,\;e\;,\;g\;$,
we get the general solution in the form,
\bea
G &=& -e^{-2a\,x}\; dx^0{}^2 +e^{-2e\,x}\;d\vec x{}^2
+\frac{1}{\Lambda^2}\;\frac{dx^2}{\sinh^2 x}+e^{-2g\,x}\;d z^2\cr
e^{2\Phi}&=& 2\, e^{2\,\Phi_0}\; e^{-\varphi\, x}\; |\sinh x|\qquad,\qquad \varphi\equiv a + p\,e + g
\eea
while the constraint equation (\ref{constraint}) reads,
\be
a{}^2 + p\,e{}^2 + g{}^2 = 1\label{consq0}
\ee
We can put the solution in a more familiar form by introducing the variables
$\rho$ and $\theta$,
\be
e^{-x} \equiv \tanh\rho\qquad,\qquad
\theta \equiv\frac{\Lambda}{2}\, z
\ee
in terms of which, after trivial re-scalings of the $x^\mu$'s,  the solution is written as,
\bea
G&=& -(\tanh \rho)^{2a}\;dx^0{}^2+(\tanh \rho)^{2e}\;d\vec x{}^2
+ {4\over \Lambda^2}\;\left(d\rho^2+(\tanh \rho)^{2g}\;d\theta^2\right)\cr
e^{2 \Phi}&=&e^{2 \Phi_0}\:\:{(\sinh\rho)^{\varphi-1}\over
(\cosh\rho)^{\varphi+1}}\qquad,\qquad \varphi\equiv a+p\,e+g\cr
1&=&a^2 + p\,e^2 + g^2\label{vacuadelta<0}
\eea
The solutions are determined, other than by $\Phi_0$, by the three parameters
$a , e, g,$ subject to the constraint in the last equation of (\ref{vacuadelta<0}).
All of them go asymptotically, i.e. for large $\rho$, to the linear
dilaton of the past subsection.
As particular solutions belonging to this three-parameter family, we have two well-known ones,
\bigskip

\noindent\underline{$a =e =0\;,g =1\:\:\:$}
\bigskip

We get the vacuum solution corresponding to the direct product of $p+1$ dimensional
Minkowski space and the cigar with scale $r_0 =\frac{2}{\Lambda}$,
\bea
G&=&\eta_{1,p}+r_0^2\;\left( d^2\rho+ \tanh^2\rho\;d\theta^2\right)\cr
e^{2\Phi}&=&e^{2 \Phi_0}\:\:{1\over \cosh^2\rho}
\eea
Here the periodicity $\theta\sim \theta + 2\,\pi$ is imposed to avoid a conical singularity at the origin $\rho=0$.
\bigskip

\noindent\underline{$a =e =0\;,g =-1\:\:\:$}
\bigskip

We get the vacuum solution corresponding to the direct product of $p+1$ dimensional
Minkowski space and the trumpet,
\bea
G &=&\eta_{1,p}+r_0^2\;\left(d^2\rho+\coth^2\rho\;d\theta^2\right)\cr
e^{2\Phi}&=&e^{2\Phi_0}\:\:{1\over\sinh^2\rho}
\eea
This solution is singular at the origin $\rho=0$.
However it is well known that the cigar and the trumpet are related by T-duality.
Both solutions correspond to exact two dimensional CFT on the world sheet,
the gauged Wess-Zumino-Witten-Novikov (WZWN) model $SL(2,\Re)/U(1)$, with vector and axial gauging respectively
\cite{Witten:1991yr}, \cite{Dijkgraaf:1991ba}.

It is interesting to note that all the solutions present a non trivial dilaton;
however for those with $\varphi\geq 1$ the string coupling $\,g_s\equiv e^\Phi\,$
is bounded everywhere, what assures us that perturbative string theory set-up is under control.
But among them, there is a two-parameter family of solutions with special features, that which
saturates the bound $\varphi=1 $.
In first term, they present the {\it same} dilaton of the cigar solution; it is also shown
that this is the only case in which the parameter $\Phi_0$ is physical, in the sense that
it is the value of the dilaton at some point (at the tip $\rho=0$).
And what is more, from the Ricci scalar of these solutions,
\be
R = \frac{\Lambda^2}{4}\; \frac{4\,\sinh^2\rho -(\varphi -1)^2 +1 -
a^2-p\,e^2-g^2}{\sinh^2\rho\,\cosh^2\rho}
= \frac{1}{r_0{}^2}\;\frac{4\,\sinh^2\rho -(\varphi -1)^2 }{\sinh^2\rho\,\cosh^2\rho}
\ee
where in the last step we have made use of the constraint in (\ref{vacuadelta<0}),
we see that they are the only regular solutions, their scalar curvature being the same as
that of the cigar.
Moreover, in contrast to the cigar solution, the other ones present necessarily a warp factor
that diverges at $\rho=0$.
For the two parameter $\varphi = 1$ family, the Ricci tensor results,
\bea
R_{00} = -\frac{2\,a}{r_0{}^2}\; \frac{1}{\cosh^2\rho}\qquad&,&\qquad
R_{IJ} = \frac{2\,e}{r_0{}^2}\; \frac{1}{\cosh^2\rho}\;\delta_{IJ}\cr
R_{p+1p+1} = \frac{2}{r_0{}^2}\; \frac{1}{\cosh^2\rho}\qquad&,&\qquad
R_{p+2p+2} = \frac{2\,g}{r_0{}^2}\; \frac{1}{\cosh^2\rho}
\eea

It would be interesting to see if some of the solutions (\ref{vacuadelta<0}) have an
exact CFT field description as the cigar and trumpet do.
We mention that in reference \cite{Alvarez:2000it} the Einstein space corresponding to the solution
$a=e=g=\frac{1}{\sqrt{p+2}}$ ($\varphi =\sqrt{p+2}$) was considered, while that the subfamily of possible vacua corresponding to
$a = e$ appears recently in reference \cite{Kuperstein:2004yk}.

\subsection{Solutions with $\Delta>0$.}

The family of solutions is given by,
\bea
G &=& -e^{-2\,a\,x}\; dx^0{}^2 +e^{-2\,e\,x}\;d\vec x{}^2
+\frac{1}{\Lambda^2}\;\frac{dx^2}{\cos^2 x}+e^{-2\,g\,x}\;dz^2\cr
e^{2\Phi}&=&2\, e^{2\,\Phi_0}\;e^{ - \varphi\, x}\;|\cos x|\qquad,\qquad \varphi \equiv a + p\,e + g
\eea
However the constraint (\ref{consq0}) is now replaced by,
\be
a{}^2 + p\,e{}^2 + g{}^2 = -1
\ee
We conclude that there exist no solutions in the branch $\Delta>0$.
\bigskip

We have exhausted all the possible uncharged solutions.
Let us go to the case of charged solutions.

\section{NSNS charged solutions, $Q_p\neq 0\;,\;b_p =-2$.}
\cleqn

These solutions, from the string theory point of view, are relevant just for $p=1$,
with the identification of the gauge field $A_2$ with the usual Kalb-Ramond two-form
gauge field $B$, under which a fundamental string can be charged.
We will maintain $p$ free however, fixing it in the analysis of relevant cases.

The equations for the $f_i$'s (\ref{system}) get decoupled,
\bea
(\ln f_1)''&=&2\,\Lambda^2\; f_1 \cr
(\ln f_2)''&=& (p+1)\,Q_p{}^2\;f_2
\eea
The solutions to each equation were worked out in the past section; we have from (\ref{f1sn})
and the positivity of $f_i$'s,
\be
f_1(x) =
\left\{\begin{array}{lcl}
{-\Delta_1 \over \Lambda^2} \:{1\over \sinh^2\left(\sqrt{-\Delta_1}\,(x-x_1)\right)}
\;\;&,&\;\;\Delta_1<0\cr
\frac{1}{\Lambda^2\,(x-x_1)^2}\;\;&,&\;\;\Delta_1=0\cr
{\Delta_1\over \Lambda^2}\:{1\over \cos^2\left(\sqrt{\Delta_1}\,(x-x_1)\right)}
\;\;&,&\;\;\Delta_1> 0
\end{array}\right.
\ee
\be
f_2(x) = \left\{\begin{array}{lcl}
{-2\,\Delta_2 \over (p+1)\,Q_p{}^2} \:{1\over \sinh^2\left(\sqrt{-\Delta_2}\,(x-x_2)\right)}
\;\;&,&\;\;\Delta_2<0\cr
\frac{2}{(p+1)\,Q_p{}^2}\,\frac{1}{(x-x_2)^2}\;\;&,&\;\;\Delta_2=0\cr
{2\,\Delta_2\over (p+1)\,Q_p{}^2}\:{1\over \cos^2\left(\sqrt{\Delta_2}\,(x-x_2)\right)}
\;\;&,&\;\;\Delta_2> 0
\end{array}\right.
\ee
where $\Delta_i , x_i,$ are arbitrary constants.
The fields are expressed as in (\ref{gralsolution}),
\bea
G &=&e^{-\frac{2\alpha}{p+1}\,x}\; f_2{}^\frac{1}{p+1}\;
\left(-e^{2\alpha\,x}\;dx^0{}^2 +\;d\vec x{}^2\right)
+f_1(x)\;dx^2+e^{2\gamma\,x}\;dz^2\cr
e^{4\Phi}&=&V_z{}^2\,e^{2\gamma\,x}\;\frac{f_2}{f_1}\cr
F_{p+2} &=& Q_p\, f_2(x)\; dx\wedge dx^0\wedge\dots\wedge dx^p
\eea
and then, they are determined by the choice of $f_1$ and $f_2$, and the imposition
of the constraint (\ref{constraint}), that reads,
\be
\frac{\Delta_2}{p+1} - \Delta_1 = \frac{p}{p+1}\;\alpha{}^2 + \gamma{}^2
\ee
The solutions depend on the parameters $\alpha, \gamma, \Delta_i , x_i$.
We collect them below
\footnote{
We alert the reader that the expressions of the fields are got after redefinitions, etc., as
in Section $3$, and therefore the $x$ coordinate, in general, is not that defined in
(\ref{xdef}).
}
.
\bigskip

\begin{enumerate}
\item \underline{$\Delta_1 = \Delta_2 = 0$}
\bea
G &=& |x-x_0|^{-\frac{2}{p+1}}\;\eta_{1,p}
+\frac{1}{\Lambda^2}\;\frac{dx^2}{x^2}+ R_1{}^2\;d\theta^2\cr
e^{2\,\Phi}&=&e^{2\,\Phi_0}\; \left|1-\frac{x_0}{x}\right|^{-1}\cr
F_{p+2} &=& s(q)\;\sqrt{\frac{2}{p+1}}\;\frac{1}{(x-x_0)^2}\;
dx\wedge dx^0\wedge\dots\wedge dx^p
\eea
where $s(q) \equiv sign (q)$, and $\theta\sim\theta +2\,\pi$, being the $S^1$ radius fixed to be,
\be
R_1 = \sqrt{\frac{p+1}{2}}\;\frac{|Q_p|\,e^{2\Phi_0}}{2\,\pi\,\Lambda}\qquad.\label{rs1}
\ee

According to the value of $x_0$, three possibilities are present.

\bigskip

If $x_0 = 0$, with $x= (r_0\,u)^{-p-1}$ the solution reads,
\bea
G &=& r_0{}^2\; \left( \frac{d u^2}{u^2} + u^2\;\eta_{1,p}\right) +R_1{}^2\; d\theta^2
\qquad,\qquad r_0\equiv\frac{p+1}{\Lambda}\cr
\Phi&=& \Phi_0\cr
F_{p+2} &=& -s(q)\sqrt{2\,(p+1)}\;r_0{}^{p+1}\; u^p\;du\wedge dx^0\wedge\dots\wedge dx^p
\label{adss1}
\eea

\bigskip

It is a $AdS_{1,p+1}|_{r_0}\times S^1|_{R_1}$ space, with a constant dilaton.
This background (for $p=1$) is well-known, it is the exact (super) CFT defined by the
$Sl(2,\Re)_{-k}\times U(1)|_{R_1}$ WZWN model,
with the level given by $(k) k-2 = \frac{4}{\alpha'\,\Lambda^2}$ \cite{Maldacena:2000hw}.

If $x_0<0$, after re-scaling $x^\mu, x\rightarrow |x_0|^\frac{1}{p+1}\,
x^\mu\, ,|x_0|\,x$, and introducing the radial variable $r$,
\be
\frac{1}{1 + x} = 1- \left(\frac{r_h}{r}\right)^{p+1} = f(r) ,\qquad,\qquad r_h <r<\infty
\ee

\bigskip

the solution takes the form,

\bea
G &=& f(r)^{\frac{2}{p+1}}\;\eta_{1,p} + r_0{}^2\; \frac{dr^2}{r^2\,f(r)^2}
+ R_1{}^2\; d\theta^2\cr
e^{2\Phi}&=& e^{2\Phi_0}\;\left(\frac{r_h}{r}\right)^{p+1}\cr
F_{p+2} &=& s(q)\;\sqrt{2\,(p+1)}\;\frac{r_h{}^{p+1}}{r^{p+2}}\;
\;dr\wedge dx^0\wedge\dots\wedge dx^p\label{f1ld}
%G &=& \left(1 + \left(\frac{u_0}{u}\right)^{p+1}\right)^{-\frac{2}{p+1}}\;\eta_{1,p}
%r_0{}^2\;\left(\frac{d u^2}{u^2} + R_1{}^2\;d\theta^2\right)\cr
%e^{2\Phi}&=& e^{2\Phi_0}\;\left(1 + \left(\frac{u}{u_0}\right)^{p+1}\right)^{-1}\cr
%F_{p+2} &=& -s(q)\;\sqrt{2\,(p+1)}\;u_0{}^{-p-1}\;
%\left(1 + \left(\frac{u}{u_0}\right)^{p+1}\right)^{-2}\;U^p\;dU\wedge dx^0\wedge\dots\wedge dx^p
\eea
It is easy to see that this solution interpolates between the linear dilaton vacuum
of Section $4.2$ (for large $r\gg r_h$) and the
$AdS_{1,p+1}|_{r_0}\times S^1|_{R_1}$ space of
(\ref{adss1}) (for $r\rightarrow r_h{}^+$).
We are tempted to identify it (for $p$=1) with a fundamental string ($F1$) embedded in the linear
dilaton vacuum, the $AdS$ space being the near horizon limit that washes off the vacuum region,
like usually happens in critical brane solutions
\footnote{
This near horizon limit can be formally done by introducing the variable $u$,
\be
(r_0\, u)^{p+1} = \left(\frac{r}{r_h}\right)^{p+1} - 1
\ee
and by taking the low energy limit $r_0\rightarrow 0$, at fixed $u$.
A related solution is considered in reference \cite{Alvarez:2001ta}.
}.
The scalar curvature is displayed in Figure $1$.

\begin{figure}[!ht]
\centering
\includegraphics[scale=0.5,angle=-90]{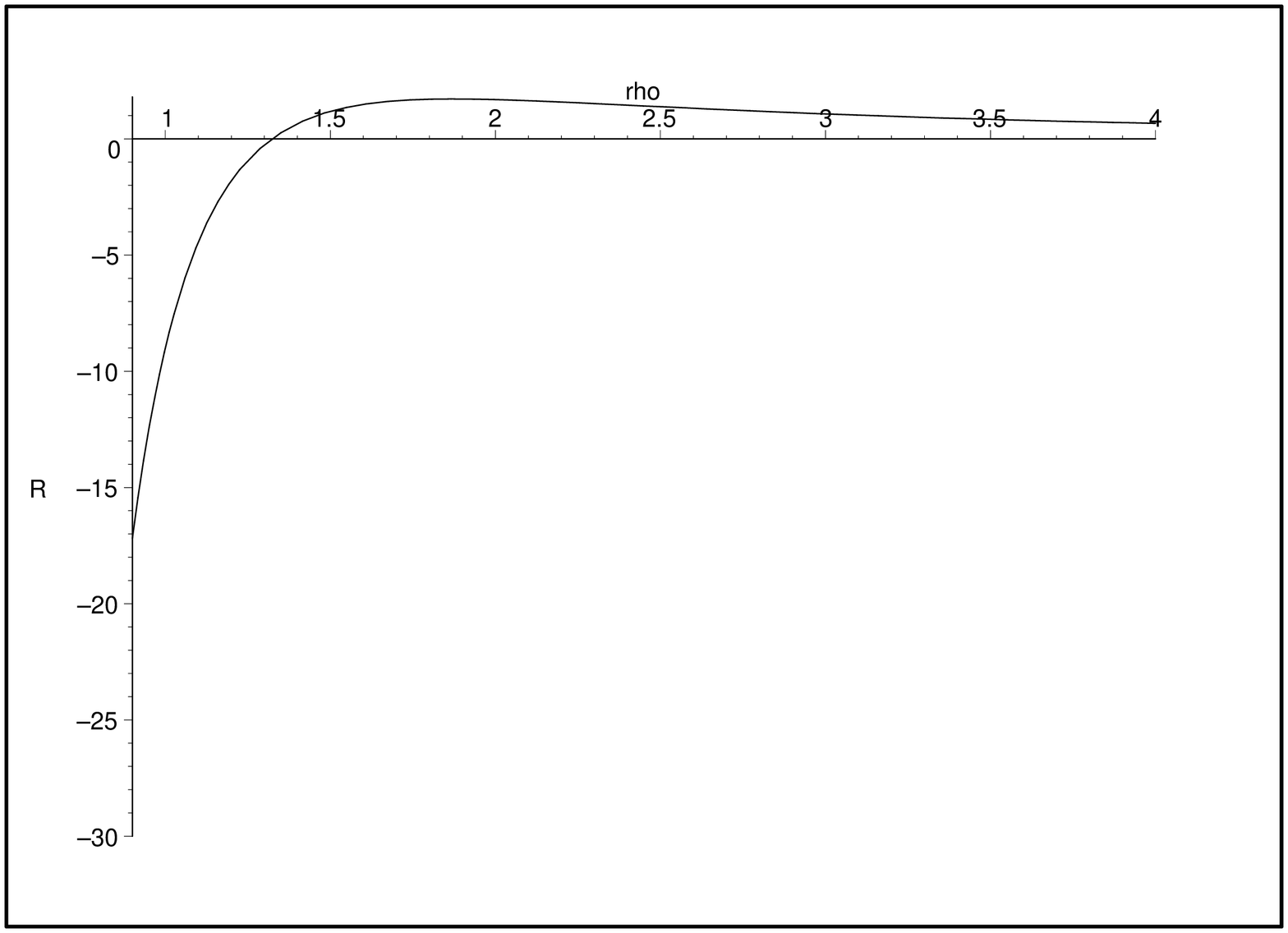}
\caption{The curve shows $\alpha'\,R$ as a function of $\rho=r/r_h$,
where $R$ is the scalar curvature corresponding to (\ref{f1ld}) and $1< \rho < \infty$.}
\label{fig1}
\end{figure}

\bigskip

Finally, if $x_0>0$ we get,

\bea
G &=& r_0{}^2\;\left(u^2\;\eta_{1,p} +\left(1 + \left(r_0\,u\right)^{p+1}\right)^{-2}\;
\frac{d u^2}{u^2}\right) + R_1{}^2\;d\theta^2 \cr
e^{2\Phi}&=& e^{2\Phi_0}\;\left(1 + \left(r_0\,u\right)^{p+1}\right)\cr
F_{p+2} &=& -s(q)\;\sqrt{2\,(p+1)}\;r_0{}^{p+1}\;\;u^p\;du\wedge dx^0\wedge\dots\wedge dx^p
\label{5.10}
\eea
which goes to $AdS_{1,p+1}|_{r_0}\times S^1|_{R_1}$ for $r_0\,u\rightarrow 0$,
but it is singular in the large $u$ region $r_0\,u\rightarrow \infty$.
The scalar curvature is displayed in Figure $2$.

\bigskip

\begin{figure}[!ht]
\centering
\includegraphics[scale=0.5,angle=-90]{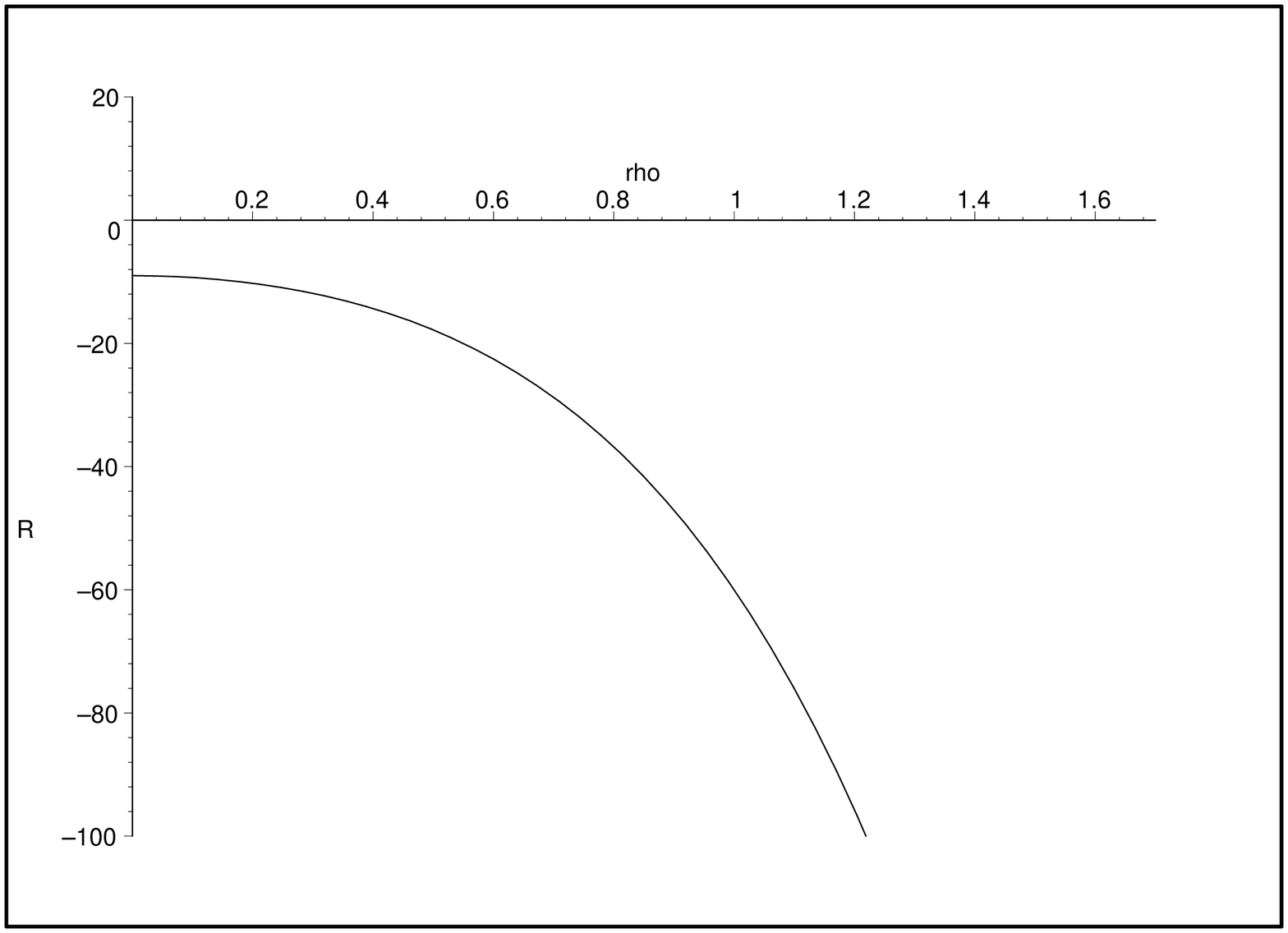}
\caption{ The curve shows $\alpha'\,R$ as a function of $\rho=r_0\,u$,
where $R$ is the scalar curvature corresponding to (\ref{5.10}) and $0< \rho < \infty$.}
\label{fig2}
\end{figure}

\item \underline{$\Delta_1 <0\;,\; \Delta_2 = 0$}
\bea
G &=& \frac{e^{-\frac{2\,a}{p+1}\,x}}{|x-x_0|^\frac{2}{p+1}}\;
\left(-e^{2\,a\, x}\, dx^0{}^2 + d{\bf x}^2\right) + \frac{1}{\Lambda^2}\,
\frac{dx^2}{\sinh^2x} + R_1{}^2\,e^{2\,g\, x}\,d\theta^2\cr
e^{2\,\Phi}&=& e^{2\,\Phi_0}\, e^{g\, x}\; \left|\frac{\sinh x}{x-x_0}\right|\cr
F_{p+2} &=& s(q)\;\sqrt{\frac{2}{p+1}}\; \frac{1}{(x-x_0)^2}\;
dx\wedge dx^0\wedge\dots\wedge dx^p\cr
1&=&\frac{p}{p+1}\,a^2 + g^2
\eea
where $R_1$ is given in (\ref{rs1}).
It is a two-parameter family of solutions, with three branches depending on the sign of $x_0$.
Let us focus on the solutions with $a=0, g=-1$, that have a bounded string coupling at large $x$.
A convenient change of variable is $e^{-x} = \tanh\rho$.

If $x_0=0$, we get,
\bea
G&=&|\ln\tanh\rho|^{-\frac{2}{p+1}}\;\eta_{1,p} +
\frac{4}{\Lambda^2}\;d\rho^2 + R_1{}^2\;\tanh^2\rho\;d\theta^2\cr
e^{-2\Phi}&=& 2\;e^{-2\Phi_0}\; \cosh^2\rho\;|\ln\tanh\rho|\cr
F_{p+2} &=&
-s(q)\;\sqrt{\frac{2}{p+1}}\;\frac{1}{\sinh\rho\;\cosh\rho\;(\ln\tanh\rho)^2}
\;d\rho\wedge dx^0\wedge\dots\wedge dx^p\cr \label{5.12}
& & \eea

The solution results, for large $\rho$, asymptotic to the
$AdS_{1,p+1}|_\frac{p+1}{\Lambda}\times S^1{}|_{R_1}$ solution of (\ref{adss1}),
but singular at $\rho=0$, where the world volume of the
brane shrinks to zero size, and the $\Re^2$ transverse space
presents a conical singularity, unless the relation
$R_1=\frac{2}{\Lambda}$ (and therefore, $\,|Q_p|\,e^{2\Phi_0}\sim
1$) be imposed, in which case the transverse space is just the cigar.
The scalar curvature is displayed in Figure $3$.

\begin{figure}[!ht]
\centering
\includegraphics[scale=0.5,angle=-90]{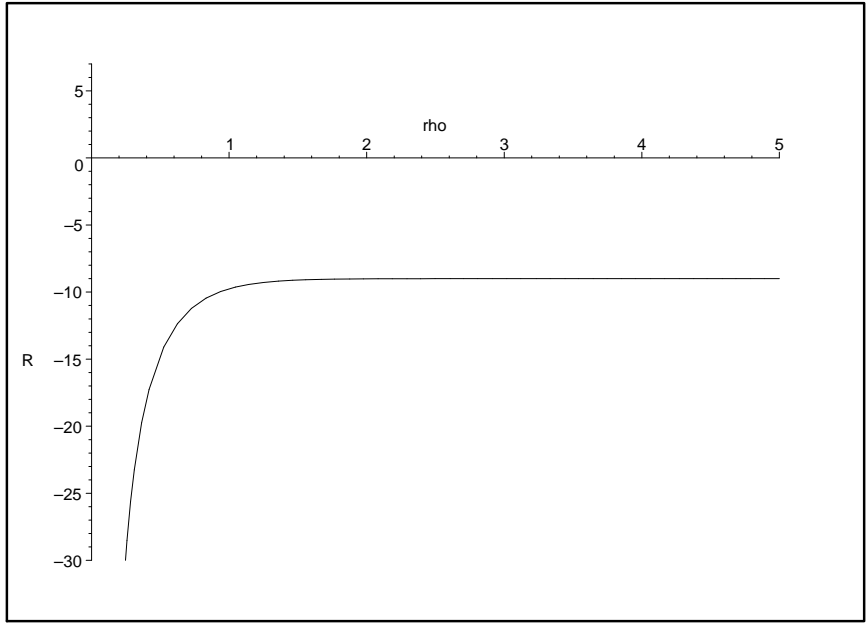}
\caption{The curve shows $\alpha'\,R$ as a function of $\rho$,
where $R$ is the scalar curvature corresponding to (\ref{5.12}),
$r_0=2/\Lambda$, and $0< \rho < \infty$.} \label{fig3}
\end{figure}

If $x_0\neq 0$, after a re-scaling $x^\mu\rightarrow
|x_0|^\frac{1}{p+1}\,x^\mu$, and the redefinition
$e^{2\Phi_0}\rightarrow 2\,|x_0|\,e^{2\Phi_0}$, the solution
reads,
\bea
G &=&|U(\rho)|^{-\frac{2}{p+1}}\; \;\eta_{1,p} +
\frac{4}{\Lambda^2}\, \left(d\rho^2 +\tanh^2\rho\;d\theta^2\right)\cr
e^{-2\Phi}&=& e^{-2\Phi_0}\,\cosh^2\rho\; |U(\rho)|\cr F_{p+2}
&=& s(q\,x_0)\;\sqrt{\frac{2}{p+1}}\; d U(\rho)^{-1}\wedge
dx^0\wedge\dots\wedge dx^p \label{eq<=0}
\eea
where,
\footnote{
The choice of such $|x_0|$ leads to the cigar as transverse space.
}
\be U(\rho) = 1 + \frac{1}{x_0}\,\ln\tanh\rho\qquad,\qquad |x_0|\equiv
\sqrt{\frac{2}{p+1}}\;\frac{2\pi}{|Q_p|}\,e^{-2\Phi_0}<0\label{f1cigar}
\ee
When $x_0<0$, we identify (for $p=1$) the solution as a fundamental non critical string embedded
in the cigar vacuum.
On the other hand, solutions with $x_0 >0$ presents a singularity at
$\rho=\rho_0, \tanh\rho_0 = e^{-\frac{2\pi}{|Q_p|e^{2\Phi_0}}}$.
These solutions were recently discovered in reference \cite{Lugo:2005yf}.
The scalar curvature of them is displayed in Figure $4$.

\begin{figure}[!ht]
\centering
\includegraphics[scale=0.5,angle=-90]{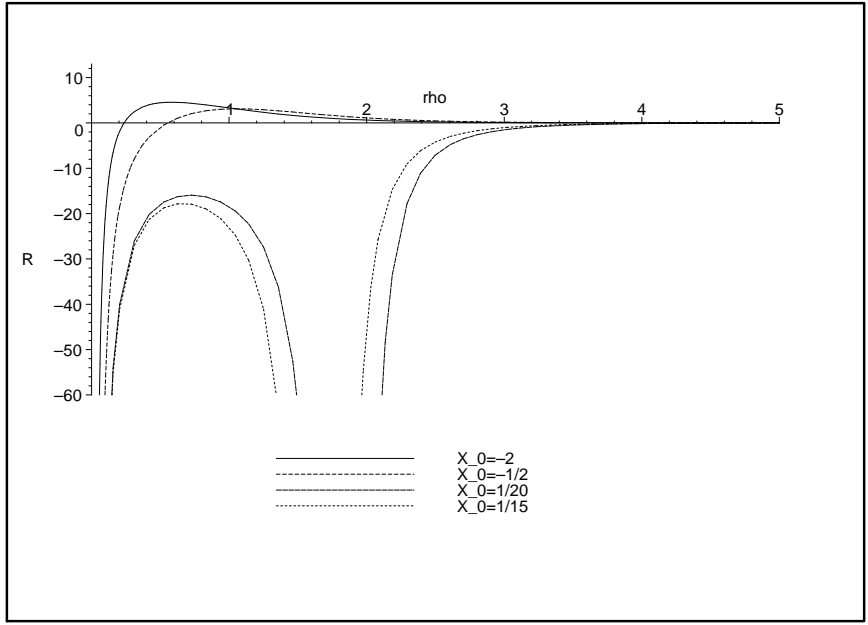}
\caption{The curves show  $\alpha'\,R$ as a function of $\rho$ for different values of $x_0$,
where $R$ is the scalar curvature corresponding to (\ref{eq<=0}).}
\label{fig4}
\end{figure}
%\pagebreak

\item \underline{$\Delta_1 <0\;,\; \Delta_2 < 0$}
\bea G &=& \left(\frac{e^{-a\,x}}{|\sinh\left(k\,(x-x_0)\right)|}\right)^{\frac{2}{p+1}}
\left(-e^{2\, a\, x}dx^0{}^2 +d\vec x{}^2\right)
+ \frac{1}{\Lambda^2}\, \frac{dx^2}{\sinh^2x}+R_k{}^2\,e^{2\,g\,x}\;d\theta^2\cr
e^{2\Phi}&=&e^{2\Phi_0}\,\frac{e^{g\,x}\; |\sinh x|}{|\sinh (k\,(x-x_0))|}\cr
F_{p+2}&=&s(q)\;\sqrt{\frac{2}{p+1}}\;\frac{k}{\sinh^2
(k\,(x-x_0))}\; dx\wedge dx^0\wedge\dots\wedge dx^p\cr
1&=&\frac{p}{p+1}\;a^2 +g^2 + \frac{k^2}{p+1}\qquad,\qquad0<k \leq \sqrt{p+1}\label{nsns3}
\eea
The family of solutions depends, other than on $\Phi_0$ and $x_0$,
on the parameters $a, g, k$, obeying the constraint in (\ref{nsns3}),
while that the $S^1$ radius is given by,
\be
R_k=\sqrt{\frac{p+1}{2}}\;\frac{|Q_p|\,e^{2\Phi_0}}{2\,\pi\,k\,\Lambda}\qquad.
\label{Rk}
\ee
The solutions with $g +1-k\leq 0$ are special in the sense that, like in Section $3.2$,
they have a string coupling bounded everywhere; furthermore, from the scalar curvature
\bea
\frac{1}{\Lambda^{2}}\,R &=& 1 - \frac{5}{2}\;\left(\frac{k\,\sinh x}{\sinh(k(x-x_0))}\right)^2 -
\left( g\,\sinh x +\cosh x  - k\,\frac{\sinh x}{\tanh(k(x-x_0))} \right)^2\cr
& &\label{eq<<0}
\eea
it follows that, if the condition $k-1 = g \geq 0$ holds, then not only the string coupling
is bounded, but  the curvature is not singular at $x=\infty$; we will focus on these subfamilies.

%An interesting particular case is $k=1$,$x_0=0$; we get a constant dilaton solution, that in the variable
%$\tanh\rho=e^{-x}$ reads,\beaG &=& |\sinh(2\rho)|^{\frac{2}{p+1}}\; \eta_{1,p} +
%\frac{4}{\Lambda^2}\, d\rho^2 + R_1{}^2\;d\theta^2\cr\Phi&=& \Phi_0\cr
%F_{p+2} &=& -s(q)\;\sqrt{\frac{2}{p+1}}\;2\,\sinh(2\rho)\;d\rho\wedge dx^0\wedge\dots\wedge dx^p
%\label{eq<<0}\eea
If $x_0 = 0$, all the family results asymptotic to
$AdS_{1,p+1}|_\frac{p+1}{\Lambda}\times S^1|_{R_k}$ when $x\rightarrow 0$.
In our string context, this means that the two-parameter subfamily with $g=k-1\geq 0$ and
$k = \frac{2}{3}\,(1+\sqrt{1-\frac{3}{4}a^2})$ determined by the constraint, presents regular
curvature everywhere and is asymptotic to $AdS_{1,2}|_\frac{2}{\Lambda}\times S^1|_{R_k}$.
%The scalar curvature is displayed in Figure $5$.
%We notice that, after various Wick's rotations, $t\equiv -i\,\frac{2}{\Lambda}\,\rho,$ etc.,
%the solution can be interpreted as a Robertson-Walker universe times $S^1|_{R_1}$, with a
%$\frac{2\,\pi}{\Lambda}$-periodic cosmological scale
%$a(t)= |\sin(\Lambda t)|^\frac{1}{p+1}$
%\footnote{
%Cosmological solutions in string theory are considered, for example,
%\cite{Antoniadis:1990uu}, \cite{Tseytlin:1991xk}, \cite{Bergshoeff:2005bt}.}.

If $x_0>0$, a singularity at $x=x_0$ develops and the solutions are not extendible to $x=0$.
On the other hand, if $x_0<0$, the solution results asymptotic to the linear dilaton one
when $x\rightarrow 0$.
The scalar curvature in the several cases considered above is displayed in Figure 5.

\begin{figure}[ht!]
\centering\includegraphics[scale=0.5,angle=-90]{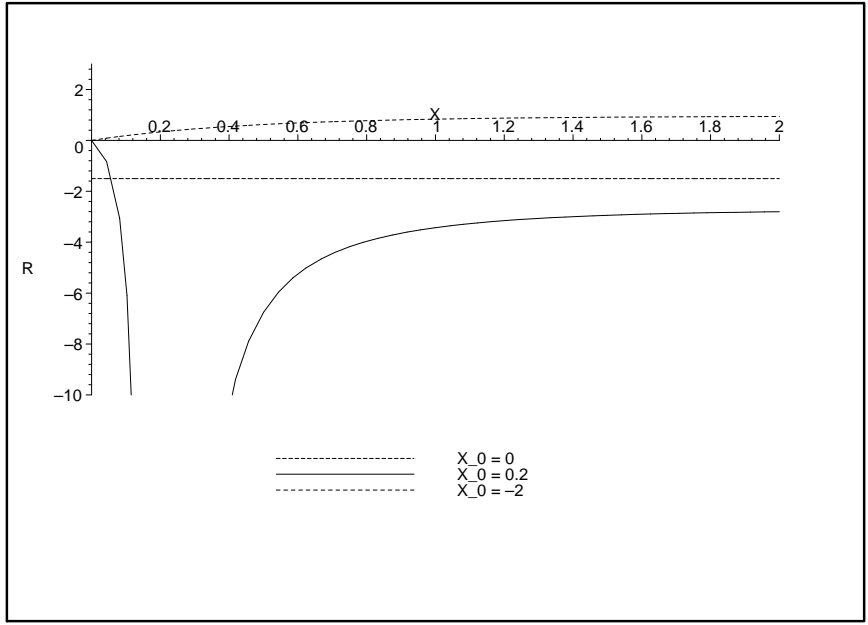}
\caption{The curve shows $\alpha'\,R$ as a function of $x$ for different values of $x_0$,
and for $g=0\,, a=k=1$; $R$ is the scalar curvature (\ref{eq<<0}).}\label{fig5}
\end{figure}

\bigskip

\item \underline{$\Delta_1 = 0\;,\; \Delta_2 > 0$}.
\bea G &=& \left(\frac{e^{-a\, x}}{|\cos(x-x_0)|}\right)^{\frac{2}{p+1}}\;
\left(-e^{2\,a\, x}\,dx^0{}^2 + d{\vec x}^2 \right)
+\frac{1}{\Lambda^2}\;\frac{dx^2}{x^2}+ R_1{}^2\; e^{2g\,x}\,d\theta^2\cr
e^{2\Phi}&=& e^{2\Phi_0}\;\frac{e^{g\,x}\,|x|}{|\cos(x-x_0)|}\cr
F_{p+2} &=&s(q)\;\sqrt{\frac{2}{p+1}}\;\frac{1}{\cos^2(x-x_0)}\;
dx\wedge dx^0\wedge\dots\wedge dx^p\cr
\frac{1}{p+1}&=&\frac{p}{p+1}\,a^2 + g^2\label{eq=>0}
\eea
where $R_1$ is given in (\ref{Rk}).
All the members of the family have an unbounded string coupling, and are
periodically singular.
We do not know how to make sense of them.
The scalar curvature is displayed in Figure $6$ for a particular value of $a, g$.

\begin{figure}[!ht]
\centering\includegraphics[scale=0.5,angle=-90]{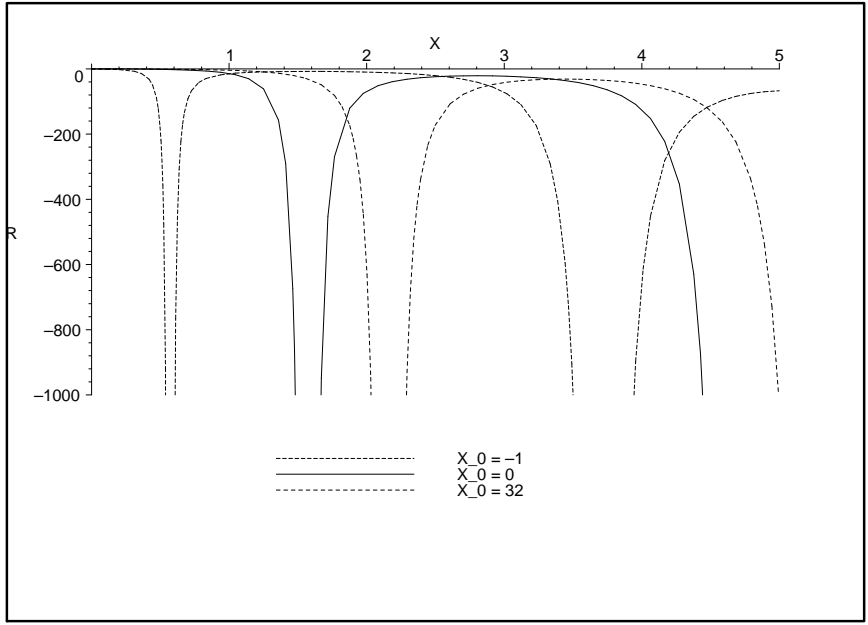}
\caption{The curves show $\alpha'\,R$ as a function of $x$ for different values of $x_0$,
where $R$ is the scalar curvature corresponding to (\ref{eq=>0}) for $a=1$ and $g=0$.} \label{fig6}
\end{figure}

\item \underline{$\Delta_1 < 0\;,\; \Delta_2 > 0$}.
\bea G &=&\left(\frac{e^{-a\, x}}{|\cos\left(k\,(x-x_0)\right)|}\right)^{\frac{2}{p+1}}
\left(-e^{2\,a\,x} dx^0{}^2 + d\vec x{}^2\right)
+\frac{1}{\Lambda^2}\,\frac{dx^2}{\sinh^2x}+R_k{}^2 e^{2g\,x} d\theta^2\cr
e^{2\Phi}&=&e^{2\Phi_0}\;\frac{e^{g\,x}\;|\sinh x|}{|\cos\left(k\,(x-x_0)\right)|}\cr
F_{p+2} &=&s(q)\;\sqrt{\frac{2}{p+1}}\;\frac{k}{\cos^2\left(k\,(x-x_0)\right)}\;
dx\wedge dx^0\wedge\dots\wedge dx^p\cr
\frac{k^2}{p+1} &=& \frac{p}{p+1}\;a^2 +g^2 -1\qquad,\qquad k>0 \label{eq<>0}
\eea
As for the solutions $4.$, they seem to have (physically) no sense.
The scalar curvature is displayed in Figure $7$.

\begin{figure}[!ht]
\centering
\includegraphics[scale=0.5,angle=-90]{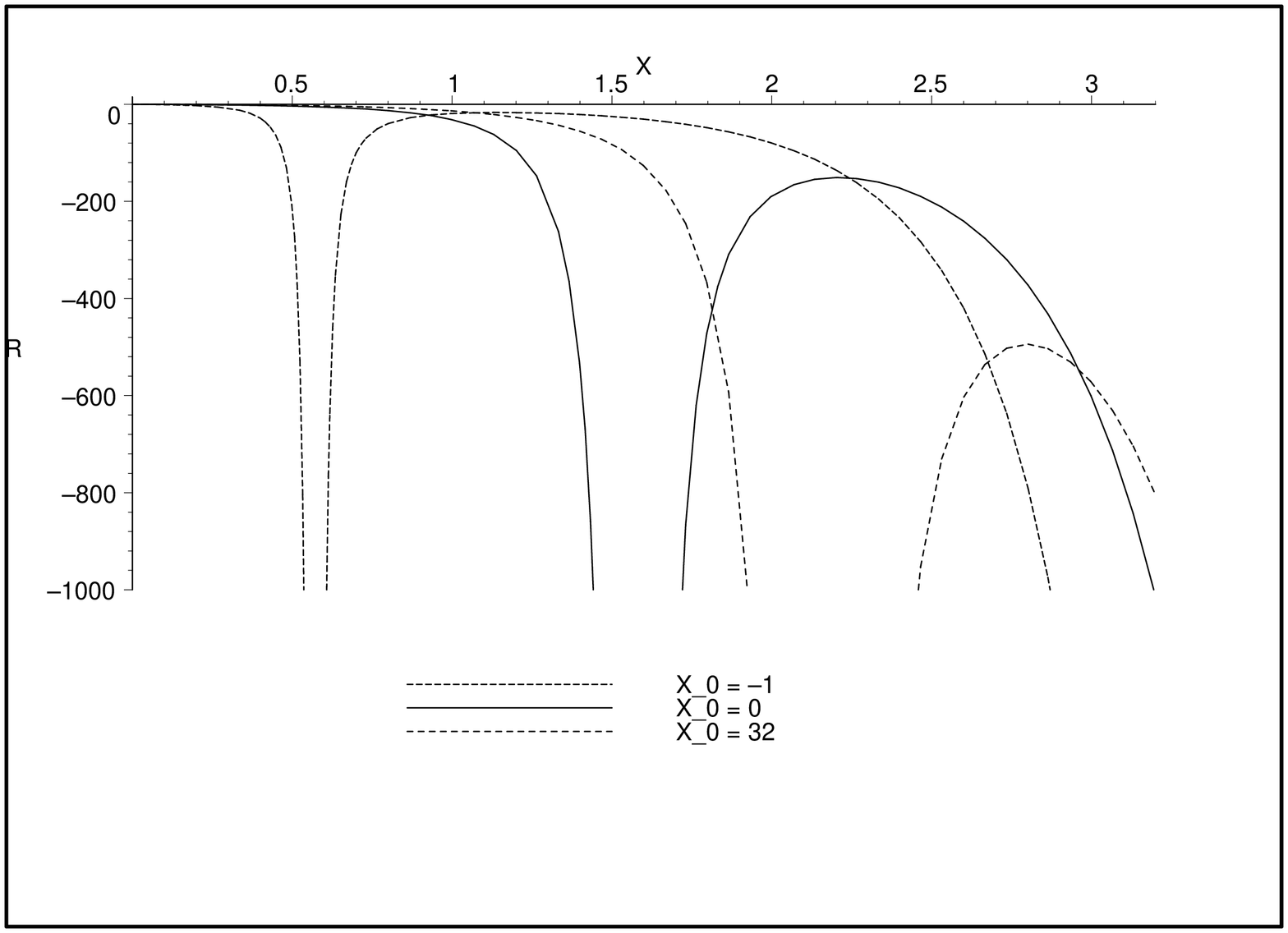}
\caption{The curves show $\alpha'\,R$ as a function of $x$ for different values of  $x_0$ and $k=a=g=1$,
where $R$ is the scalar curvature corresponding to (\ref{eq<>0})}.\label{fig7}
\end{figure}

\item \underline{$\Delta_1 > 0\;,\; \Delta_2 > 0$}.
\bea
G &=&\left(\frac{e^{-\,a\,x}}{|\cos\left(k\,(x-x_0)\right)|}\right)^{\frac{2}{p+1}}
\left(-e^{2\,a\,x}\,dx^0{}^2 +d\vec x{}^2\right)
+\frac{1}{\Lambda^2}\,\frac{dx^2}{\cos^2x}+R_k{}^2\;e^{2g\,x}\,d\theta^2\cr
e^{2\Phi}&=&e^{2\Phi_0}\;\frac{e^{g\,x}\;|\cos x|}{|\cos\left(k\,(x-x_0)\right)|}\cr
F_{p+2} &=&s(q)\;\sqrt{\frac{2}{p+1}}\;\frac{k}{\cos^2\left(k\,(x-x_0)\right)}\;
dx\wedge dx^0\wedge\dots\wedge dx^p\cr
\frac{k^2}{p+1} &=& \frac{p}{p+1}\;a^2 + g^2 +1\qquad,\qquad k\geq \sqrt{p+1}
\eea
The remarks made for the families $4.$ and $5.$ also apply, the behavior of the scalar curvature
being similar, and we do not show it.
%except for the solution with $k=1$, $x_0=0$, which presents a constant dilaton;
%after the change of variable, $\cos x = \frac{1}{\cosh Y}$, it reads,
%\bea G&=&|\cosh Y|^\frac{2}{p+1}\;\eta_{1,p}+\frac{1}{\Lambda^2}\,dY^2 +R_1{}^2\;d\theta^2\cr
%\Phi&=& \Phi_0\crF_{p+2} &=& s(q)\;\sqrt{\frac{2}{p+2}}\;\cosh Y\;
%dY\wedge dx^0\wedge\dots\wedge dx^p \label{nsnscosmo}\eea
%The solution is regular, and asymptotic (for large $Y$) to $AdS|_{\frac{p+1}{\Lambda}}\times S^1|_{R_1}$.
%The scalar curvature is displayed in Figure $8$. As it happens with the solution (\ref{eq<<0}), it can be interpreted,
%after a double Wick rotation $Y\rightarrow i\Lambda\,t\,,\, x^0\rightarrow -i\,x^{p+1}\,$,
%as a Robertson-Walker universe times $S^1|_{R_1}$, with a $\frac{2\,\pi}{\Lambda}$-periodic cosmological scale $a(t)= |\cos(\Lambda t)|^\frac{1}{p+1}$.
\end{enumerate}

\section{RR charged solutions, $b_p = 0$.}
\cleqn

Instead of working with arbitrary $b_p \neq -2$, we specialize to
$b_p=0$, the known (in string frame) decoupling of the dilaton to
the R-R field $A_{p+1}$.
From (\ref{gralsolution}) we have,
\bea
G&=& \tilde A^2\;\left(-e^{2\alpha\,x}\; dx^0{}^2 + {d\vec x}^2\right)
+ f_1(x)\;dx^2 + e^{2\gamma\,x}\;\tilde
A^{-2}\; dz^2\cr
e^{2(p+2)\Phi} &=&V_z{}^{2p}\;
e^{2\left(\alpha+(p+1)\gamma\right)\,x}\;\frac{f_2{}^p
}{f_1{}^{p+1}}\cr
F_{p+2} &=& V_z\, Q_p \,f_2(x)\;dx\wedge dx^0\wedge\dots\wedge dx^p
\eea
where
\be \tilde A^{2(p+2)} = V_z{}^4\, e^{2\,(-\alpha+\gamma)\,x}\;\frac{f_2{}^2}{f_1}
\ee
The constraint (\ref{constraint}) becomes,
\bea
-\frac{1}{2}\,(\ln f_1)''+\frac{1}{4}\,(\ln f_1)^{'\:2} &=& \left((\ln\tilde A)' +
\alpha\right)^2 +p\,(\ln\tilde A)'{}^2 +\left((\ln\tilde A)'
-\gamma\right)^2 -\frac{Q_p{}^2}{4}\; f_2\cr
2\,(p+2)\,(\ln\tilde A)'
&=& - 2\,\alpha + 2\,\gamma + \left(
\ln\frac{f_2{}^2}{f_1}\right)'\label{consrr}
\eea
In contrast to the cases treated before in Sections $3$ and $4$, we have not succeed in solving
in complete generality equations (\ref{system}).
Instead, we will make the following ans\"atz,
\be
f_1(x) = f(x)\quad,\qquad f_2(x) =
\frac{4}{p+3}\;\frac{\Lambda^2}{Q_p{}^2}\; f(x)\label{ansatz}
\ee
Equations (\ref{system}) reduce to,
\be (\ln f)''=2\;\frac{p+2}{p+3}\,\Lambda^2\;f
\ee
The possible solutions for (non negative) $f$ were analyzed in the past
Sections, equation (\ref{f1sn}).
We have three cases, however the constraint (\ref{consrr}) reads,
\be -\Delta =\alpha{}^2 +\gamma{}^2 +\frac{2}{p+1}\,\alpha\,\gamma \label{consrr1}
\ee
which rules out the possibility $\Delta>0$.

\subsection{Solution with $\Delta=0$.}

The functions $f_1 , f_2$ in (\ref{ansatz}) are,
\be f_1(x)
=\frac{p+3}{p+2}\,\frac{1}{\Lambda^2}\,
\frac{1}{(x-x_0)^2}\qquad,\qquad f_2(x) =
\frac{4}{p+2}\,\frac{1}{Q_p{}^2}\, \frac{1}{(x-x_0)^2}
\ee
and, taking into account (\ref{consrr1}), we get the following solution,
\bea
G &=& l_0{}^2\;\left(u^2\;\eta_{1,p} +\frac{d u^2}{u^2}\right) +
\frac{R_0{}^2}{(l_0\,u)^2}\;d\theta^2 \cr
e^{-2\Phi}&=&e^{-2\Phi_0}\;(l_0\,u)^{2}\cr
F_{p+2} &=& s(q)\; 2\sqrt{p+2}\;e^{-\Phi_0}\; l_0{}^{p+2}\; u^{p+1}\;du\wedge
dx^0\wedge\dots\wedge dx^p\label{delta0}
\eea
where the scale $l_0$ and the $S^1$ radius $R_0$ are,
\be l_0 = \sqrt{(p+2)(p+3)}\,\Lambda^{-1}\qquad,\qquad R_0 = \frac{|Q_p|\,
e^{\Phi_0}}{4\,\pi\,\sqrt{p+2}}\;l_0\label{l0R0}
\ee
It is an $AdS_{1,p+1}$ space with scale $l_0$, times a $S^1$ with
$u$-dependent radius, that enforces a running for the dilaton and
makes the solution not conformal. For large $u\gg l_0^{-1}$, the
$S^1$ shrinks to zero size, and we remain with $AdS_{1,p+1}|_{l_0}$;
instead for $u\ll l_0{}^{-1}$ it is the $Dp$ brane world volume that
shrinks, leaving a transverse $AdS_2$ space with the same scale
$l_0$. The Ricci tensor results, \be R_{\mu\nu} =
-\frac{p}{l_0{}^2}\,\eta_{\mu\nu}\qquad,\qquad R_{p+1,p+1}=
-\frac{p+2}{l_0{}^2}\,\qquad,\qquad R_{p+2,p+2}=\frac{p}{l_0{}^2}
\ee from where a constant scalar curvature follows, \be R =
-(p^2+p+2)\;\frac{1}{l_0{}^2} \ee The solution is regular
everywhere, but unfortunately the dilaton diverges when $u\rightarrow 0^+$.
We notice that it can be thought as the T-dual
solution (in $\theta$ direction) of $AdS_{1,p+2}$ space with a
constant dilaton (see at the end of this section). It was recently
obtained in reference \cite{Kuperstein:2004yk} as the near horizon
limit of a BPS solution which is asymptotic to the linear dilaton
background.

\subsection{Solutions with $\Delta<0$}

From (\ref{ansatz}), (\ref{f1sn}), we have in this case,
\bea
f_1(x)&=&\frac{-\Delta}{\Lambda^2}\; \frac{p+3}{p+2}\;\frac{1}{\sinh^2(\sqrt{-\Delta}\,(x-x_0)}
\cr
f_2(x)&=&\frac{-\Delta}{Q_p{}^2}\;\frac{4}{p+2}\;
\frac{1}{\sinh^2(\sqrt{-\Delta}\,(x-x_0))}
\eea
After re-scalings, the family of solutions can be written as,
\bea
G &=& |\sinh x|^{-\frac{2}{p+2}}\;\left(-e^{2a\,x}\; d^2 x^0 + e^{2e\,x}\; d\vec x{}^2\right)
+ \frac{l_0{}^2}{(p+2)^2}\;\frac{d^2 x}{\sinh^2 x}\cr
&+& R_0{}^2\; e^{2(a+p\,e)\,x}\;|2\,\sinh x|^\frac{2}{p+2}\; d\theta^2\cr
e^{2\Phi} &=& e^{2\Phi_0}\;e^{2(a+p\,e)\,x}\;|2\,\sinh x|^\frac{2}{p+2}\cr
F_{p+2} &=& s(q)\;\frac{2^\frac{p+1}{p+2}}{\sqrt{p+2}}\;e^{-\Phi_0}\;\frac{1}{\sinh^2 x}
\;dx\wedge dx^0\wedge\dots\wedge dx^p\cr
\frac{p+1}{p+2}&=&2\,a^2 + p\,(p+1)\,e^2 + 2\,p\,a\,e
\label{deltaless0}
\eea
where $R_0 , l_0,$ are given in (\ref{l0R0}).
It results illuminating the change of variable,
\be
e^{-2x} = 1 - \left(\frac{u_0}{u}\right)^{p+2}\equiv f(u)
\ee
with $u_0$ a constant, that puts the solutions, after trivial re-scalings, in the following form,
\bea
G &=& l_0{}^2\left(-\frac{u^2}{f(u)^{a-\frac{1}{p+2}}}\;dx^0{}^2 +
\frac{u^2}{f(u)^{e-\frac{1}{p+2}}}\;d\vec x{}^2+ \frac{1}{f(u)}\;\frac{d u^2}{u^2}\right)\cr
&+& R_0{}^2\;\frac{ f(u)^{-a-p\,e-\frac{1}{p+2}}}{(l_0\,u)^2}\;d\theta^2\cr
e^{-2\Phi} &=& e^{-2\Phi_0}\; f(u)^{a+p\,e+\frac{1}{p+2}}\; (l_0\,u)^2\cr
F_{p+2} &=& s(q)\; 2\,\sqrt{p+2}\; e^{-\Phi_0}\;l_0{}^{p+2}\; u^{p+1}\;
du\wedge dx^0\wedge\dots\wedge dx^p\cr
\frac{p+1}{p+2}&=&2\,a^2 + p\,(p+1)\,e^2 + 2\,p\,a\,e\label{adsbhsngral}
\eea
with $l_0$ and $R_0$ as in (\ref{l0R0}).
The $u_0 =0$ limit is just the $\Delta=0$ solution (\ref{delta0}), that
results the UV asymptotic limit  $u\gg u_0$ of all the family.
On the other hand, for $u\rightarrow u_0{}^+$ the behavior strongly depends
on the exponents $a, e$.
The scalar curvature for particular values of the parameters is displayed in Figures $8$ and $9$.

\begin{figure}[!ht]
\centering
\includegraphics[scale=1]{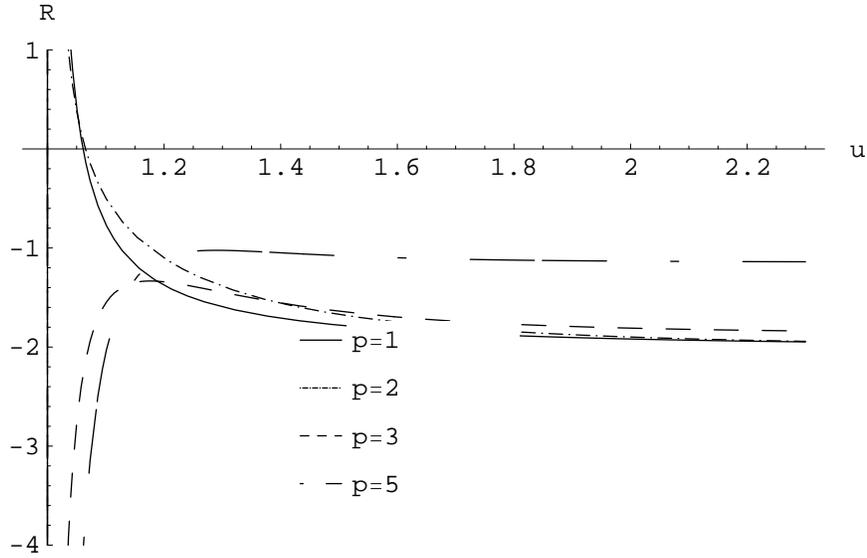}
\caption{The curves show $\alpha'\,R$ as a function of
$\tilde{u}={u \over u_0}$ for different values of $p$, where $R$ is the scalar
curvature of (\ref{adsbhsngral}) for $a=1/5$ and $e=-3/10$.}
\label{fig8}
\end{figure}

\begin{figure}[!ht]
\centering
\includegraphics[scale=1]{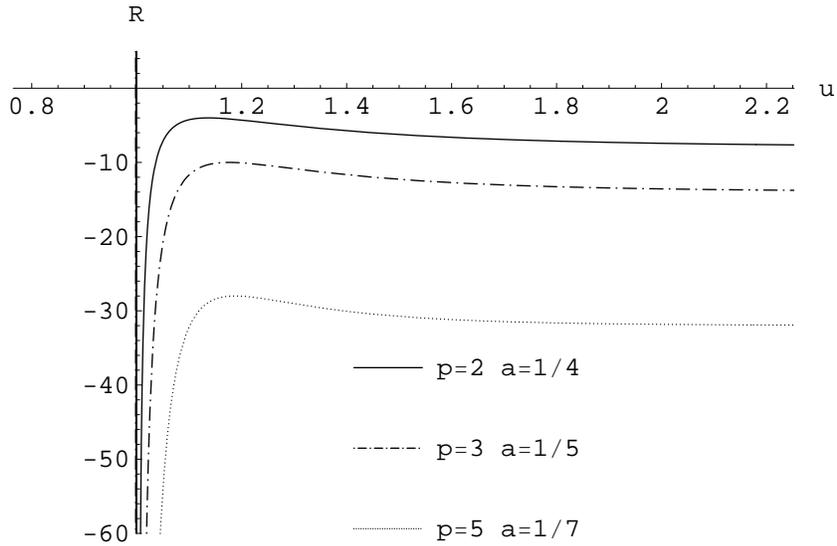}
\caption{The curves show $\alpha'\,R$ for (\ref{adsbhsngral}) as a
function of $\tilde{u}={u \over u_0}$ for different values of $p$.
The values of $a$ and $e$ are such that the solutions satisfy the Maldacena-Nu\~nez
criterion and have a bounded string coupling.} \label{fig9}
\end{figure}

Among the members of the family, those with $a + p\,e +\frac{1}{p+2}\leq 0$ have
a bounded string coupling everywhere.
And, as it happened with the family of solutions of Section $3.2$, when the inequality is
saturated, they are also regular.
In fact, the scalar curvature of (\ref{adsbhsngral}) reads,
\bea
l_0{}^2\,R &=& -\left(1+(p+2)\,(a+pe)\right)^2\;\frac{1}{f(u)} - (p-1)\,(p+2) + 2 -
2\,(p+2)^2\;(a+pe)^2\cr
&-& \left(1-(p+2)\,(a+pe)\right)^2\;f(u)
\eea
showing explicitly that the curvature is finite at $u\rightarrow u_0$ iff the condition
$a + p\,e +\frac{1}{p+2} = 0$ holds.
In this case, the following two solutions emerge.
\bigskip

\noindent\underline{Solution 1}. $a =-\frac{p+1}{p+2}\;\;,\;\;e =\frac{1}{p+2}$

\bea
G &=& l_0{}^2\; \left( -u^2\;f(u)\;dx^0{}^2+ u^2\;d^2 \vec x + \frac{1}{f(u)}\;
\frac{d u^2}{u^2}\right)+ \frac{R_0{}^2}{(l_0\,u)^2}\;d\theta^2\cr
e^{-2\Phi} &=& e^{-2\Phi_0}\;(l_0\,u)^2\cr
F_{p+2} &=& s(q)\; 2\,\sqrt{p+2}\; e^{-\Phi_0}\;l_0{}^{p+2}\; u^{p+1}\;
du\wedge dx^0\wedge\dots\wedge dx^p\label{adsbhsn1}
\eea
It is an $AdS$ like black hole solution, with a running dilaton.
Furthermore, it is not difficult to see, following the standard recipe, that in the limit
of $u\rightarrow u_0$ the euclidean space obtained from the Wick rotation $\tau\equiv i\,x^0$
is regular if we impose the periodicity,
\be
\tau \sim \tau + \beta\qquad,\qquad\beta\equiv \frac{4\,\pi}{p+2}\,\frac{1}{u_0}
\ee
This is an usual fact in a kind of solutions that we associate to field theories at
finite temperature $\beta^{-1}$ \cite{Witten:1998zw}.

\bigskip

\noindent\underline{Solution 2}.
$a =\frac{p^2+2p-1}{(p+1)\,(p+2)}\;\;,\;\;e = -\frac{p+3}{(p+1)\,(p+2)}$
\bea
G &=& l_0{}^2\; \left( -\frac{u^2}{f(u)^{\frac{p-1}{p+1}}}\;dx^0{}^2+ u^2\;
f(u)^{\frac{2}{p+1}}\;d^2 \vec x+ \frac{1}{f(u)}\;\frac{d u^2}{u^2}\right)
+\frac{R_0{}^2}{(l_0\,u)^2}\;d\theta^2\cr
e^{-2\Phi} &=& e^{-2\Phi_0}\;(l_0\,u)^2\cr
F_{p+2} &=& s(q)\; 2\,\sqrt{p+2}\; e^{-\Phi_0}\;l_0{}^{p+2}\; u^{p+1}\;
du\wedge dx^0\wedge\dots\wedge dx^p\label{adsbhsn2}
\eea
The $D1$-brane case is included in the Solution $1$ through a double Wick rotation,
$x^0\rightarrow -i\,x^1\;,\;x^1\rightarrow i\,x^0$.
We noticed that both solutions are supported by the same dilaton and $RR$ gauge fields,
and have the same scalar curvature, which is displayed in Figure $10$.

%\pagebreak

\begin{figure}[!ht]
\centering
\includegraphics[scale=0.5,angle=-90]{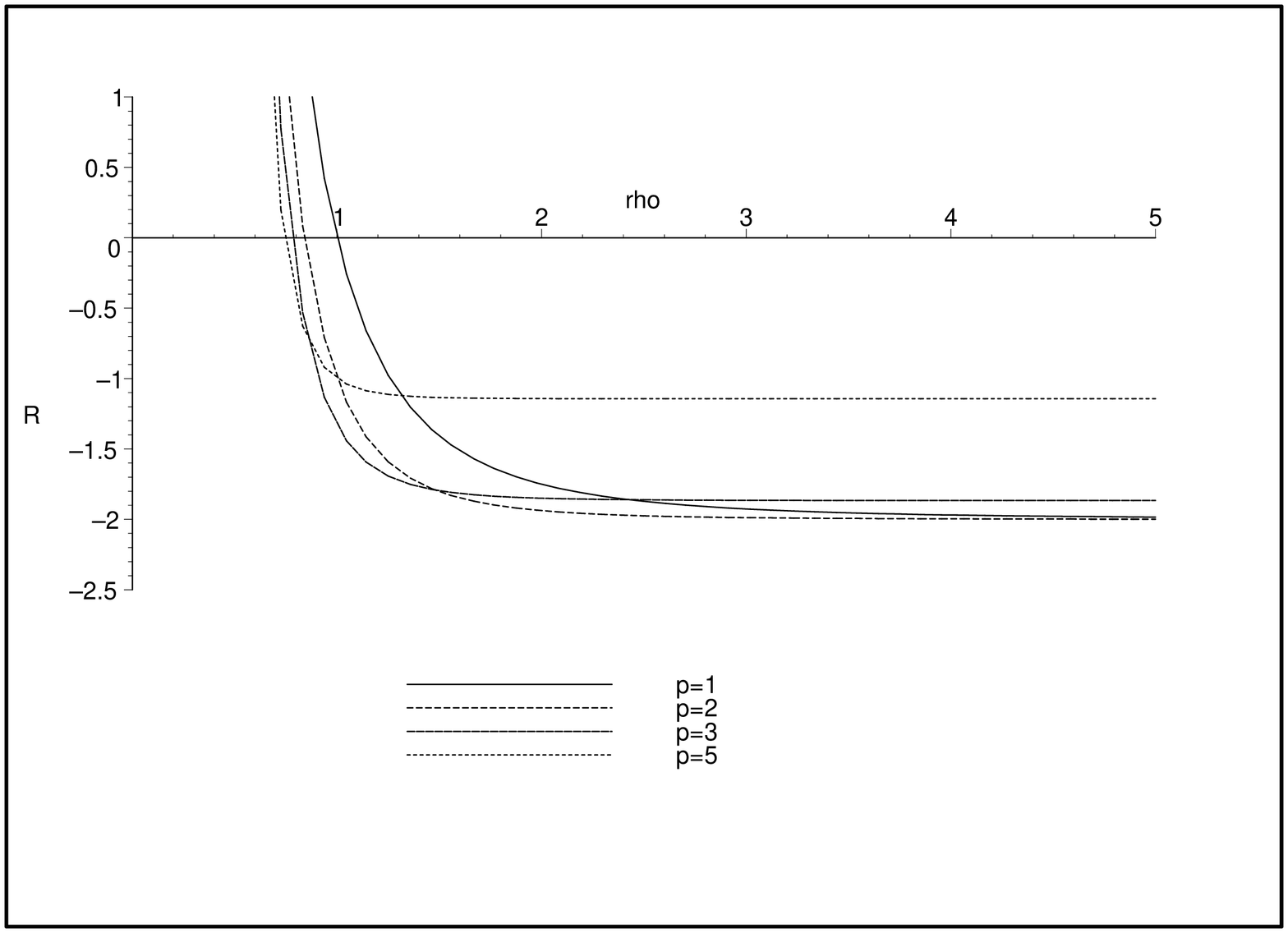}
\caption{The curves show $\alpha'\,R$ as a function of $\tilde{u}={u \over u_0}$
for different values of $p$, where $R$ is the scalar curvature of (\ref{adsbhsn1})
and (\ref{adsbhsn2}).}\label{fig10}
\end{figure}

\vspace{12cm}

\subsection{$T$-duality.}

By applying the rules for $T$-duality transformations (see i.e. \cite{Johnson:2000ch} and
references therein) along the coordinate $\theta$, we can generate from
(\ref{adsbhsngral}) another family of solutions.
The solutions obtained in this section assumed the proportionality between the functions
$f_1$ and $f_2$, equation (\ref{ansatz}); it is easy to see that this ans\"atz yields
the relation $e^{2\Phi}\propto G_{\theta\theta}$, which in turns yields T-dual solutions
with constant dilaton.
We get,
\bea
G &=& l_0{}^2\left(-\frac{u^2}{f(u)^{a-\frac{1}{p+2}}}\;dx^0{}^2 +
\frac{u^2}{f(u)^{e-\frac{1}{p+2}}}\;d\vec x{}^2+ \frac{1}{f(u)}\;\frac{d u^2}{u^2}\right.\cr
&+&\left.\left(\frac{\alpha'}{R_0}\right)^2\; f(u)^{a+p\,e+\frac{1}{p+2}}\;
u^2\;d\theta^2\right)\cr
e^{2\Phi} &=& \frac{16\,\pi^2\,\alpha'\,\Lambda^2}{(p+3)\,Q_p{}^2}\cr
F_{p+3} &=& s(q)\; 2\,\sqrt{p+2}\; e^{-\Phi_0}\;\sqrt{\alpha'}\;l_0{}^{p+2}\; u^{p+1}\;
du\wedge dx^0\wedge\dots\wedge dx^p\wedge d\theta\cr
\frac{p+1}{p+2}&=&2\,a^2 + p\,(p+1)\,e^2 + 2\,p\,a\,e\label{T-adsbhsngral}
\eea
It can be showed that this is a family of Einstein spaces with Ricci tensor,
\be
R_{mn} = - \frac{p+2}{l_0{}^2}\;G_{mn}
\ee
All the members are asymptotic at large $u\gg u_0$ to $AdS_{1,p+2}$ space
(after identifying $x^{p+1}\equiv \frac{\alpha'}{R_0}\,\theta$).
In particular, the T-dual solution to (\ref{adsbhsn1}) results the $AdS_{1,p+2}$
Schwarzchild black hole recently derived in \cite{Kuperstein:2004yk}, and used in
\cite{Kuperstein:2004yf}, \cite{Casero:2005se}, as a model of four dimensional YM theory
\footnote{
We have explicitly verified that the family (\ref{T-adsbhsngral}) is solution of the equation of motion (\ref{ecformal}).
In particular, the (dimensional) $D(p+1)$ brane charge should be identified with the original $Dp$ brane charge in the following way,
\be
Q_{p+1}\equiv \frac{Q_p}{2\,\pi\,\sqrt{\alpha'}}
\ee
Furthermore, the arbitrariness, through $\Phi_0$, in the parameter $\frac{R_0}{l_0}$ given in (\ref{l0R0})
translates here in the arbitrariness of the $x_{p+1}$ compactification radius.
}
.
\bigskip

\section{Conclusions and perspectives.}
\cleqn

In this paper, we have considered $p$-brane solutions to the
low energy gravity equations of motion of non critical type II
string theories in $D=p+3$ dimensions.
We have reduced the problem to the search for the general solution of the system given
by equations (\ref{system}), (\ref{constraint}).

Firstly, we found all the possible uncharged vacuum solutions.
Besides $p+1$ Minkowski space-time times the linear dilaton times
$S^1$, we found the three parameter family of solutions
(\ref{vacuadelta<0}), that includes the well-known $p+1$ Minkowski
space-time times the cigar, and its T-dual $p+1$ Minkowski
space-time times the trumpet. In a large region of the parameter
space the string coupling remains bounded, and there is a two
parameter subfamily that also has bounded scalar curvature.

We were also able to solve completely the problem for all the
possible NSNS charged backgrounds, which for $p=1$ represent (the
near horizon of) fundamental string solutions.
Among the large amount of found solutions, we highlight the fundamental
string solution embedded in the cigar vacuum (\ref{f1cigar}),
recently found in \cite{Lugo:2005yf}, a new solution (\ref{f1ld})
interpretable as a F1 embedded in the linear dilaton vacuum, and
a regular two-parameter family asymptotic to $AdS_{1,2}\times S^1$.

With regard to RR charged $Dp$-brane solutions, we were able to
find analytically a class of solutions obeying a particular
constraint. It includes a constant curvature solution asymptotic
in the UV to $AdS_{1,p+1}$, that shrinks to $AdS_2$ in the IR
limit; it is T-dual to $AdS_{1,p+2}$. This solution first appears
in \cite{Kuperstein:2004yk} as the near horizon of a (numerically
found) solution asymptotic to the linear dilaton vacuum.
Furthermore, we found a three parameter family of solutions
(\ref{adsbhsngral}), that presumably correspond to the near
horizon limit of a three parameter family of black $Dp$-brane
solutions, embedded in the linear dilaton vacuum, solutions that
do not obey our ans\"atz (\ref{ansatz}) in all the space-time. On
the other hand, we conjecture that there should exist $Dp$- branes
embedded in the cigar vacuum which are solutions of
(\ref{system}), that do not obey our ans\"atz in any region, and
that correspond to those constructed as boundary states in
references \cite{Ashok:2005py}, \cite{Fotopoulos:2005cn},
\cite{Murthy:2006xt} \footnote{ We thank Jan Troost for valuable
remarks about this point. } . This is subject for future work.

The condition that the solutions of (\ref{adsbhsngral}) have a
small string coupling reads,
\be
e^{2\Phi_0} \ll
f(u)^{a+p\,e+\frac{1}{p+2}}\; (l_0\,u)^2
\ee
For the regular solutions $1$ and $2$ of Section $7$, this condition reduces to,
\be
e^{2\Phi_0} \ll (l_0\,u_0)^2
\ee
and their curvature is showed in Figure $10$.
However, we can consider the region of parameters where $a+p\,e+\frac{1}{p+2}\neq 0$.
In that case, the solution will present a curvature singularity in the IR limit,
$u\rightarrow u_0{}^+$.
There is, nevertheless, nothing new in the fact that SUGRA solutions one would like to interpret
as gravity duals of some field theory in the IR, have curvature or string coupling singularities.
A known example having a curvature singularity is the Klebanov-Tseytlin solution \cite{Klebanov:2000nc},
the UV limit of the regular Klebanov-Strassler \cite{Klebanov:2002gr}, that however
is useful in discussing aspects of the dual $N=1$ gauge theory that depend on the UV
behavior of the theory, like the chiral anomaly of the R-current \cite{Klebanov:2000hb}.
More generically, singular solutions could be useful if they satisfy that $G_{00}$
does not increase as one approaches the singularity, according to the
strong form of the Maldacena-Nu\~nez criterion \cite{Maldacena:2000mw} (see also \cite{Gubser:2000nd}).
For example, the solutions in Figure $9$ satisfy this criterion.
On the other hand, if $a+p\,e+\frac{1}{p+2}> 0$, the string coupling
blows up in the IR and we should switch to its ``S-dual" solution, whatever it means.
In this respect, the connection of non critical theories with ten dimensional ones worked out
in references \cite{Giveon:1999zm}, \cite{Giveon:1999px}, \cite{Giveon:1999tq}, could be of help.

On the top of the questions that deserve further investigations, it certainly remains to explore
the possibility of using (some of) the solutions (\ref{adsbhsngral})
\footnote{
NSNS solutions asymptotic to $AdS_{3}$ spaces, in particular the regular sub-family of (\ref{nsns3}),
could also be of usefulness, in the spirit of the ${AdS_3/CFT_2}$ correspondence.
We thank Sameer Murthy for a discussion on this point.
}
as gravity dual of gauge field theories,
along the lines of references \cite{Kuperstein:2004yf}, \cite{Casero:2005se}, and in particular, to understand
the role of the exponents on the field theory side.
Probably the T-dual versions (\ref{T-adsbhsngral}) are best settled to this end,
because all of them have the same constant dilaton that, being $|Q_p| \sim N$, where $N$ is the number of $Dp$-branes
(see the discussion in footnote $3$), is given by,
\be
g_s = e^\Phi \sim \frac{1}{N}
\ee
So, for $N$ large enough, we can trust perturbative string theory for any solution of the family.
Furthermore, the computation on the gravity side of the number of degrees of freedom (``entropy") along
the lines of \cite{Kuperstein:2004yk} yields,
\be
S_{gravity} \sim \frac{N^2}{\delta^{p+1}}
\ee
where $\delta$ is an IR cutoff, which is the result we expect for a $p+2$ dimensional gauge theory with
UV cutoff $\delta^{-1}$ \cite{Susskind:1998dq}.
Finally, it can be shown \cite{lsinprep} that if the parameters obeys the constraint,
\be
a + e > \frac{2}{p+2}
\ee
then, according to the Wilson loop criterion \cite{Kinar:1998vq}, the theory is confining.
We hope to present related results in a future publication \cite{lsinprep}.
\bigskip

\noindent{\bf Acknowledgments.}

We would like to thank Juan Maldacena for a discussion in the
very early stage of these investigations, Jan Troost for useful
correspondence, and very specially  Jorge Russo for reading the
manuscript and making enlightening comments and suggestions.

%\vfill\eject

\appendix

\section{Some useful formulae.}
\cleqn

Here we collect properties related with the ans\"atz for the fields assumed in
Section $2.1$,
\bea
G &=& -A^2\; dx^0{}^2+\tilde{A}^2\;d{\vec x}{}^2 + C^2\;d\rho^2 +
\tilde C^2\;dz^2\equiv\eta_{mn}\;\omega^m\;\omega^n\cr
A_{p+1} &=& dx^0\wedge\dots\wedge dx^p\; E(\rho)\cr
\Phi &=& \Phi(\rho)\label{ansatzapen}
\eea
where $\vec x$ is a $p$ dimensional vector, and we have introduced the vielbein $\omega^n$ and
dual vector fields $e_m$, $e_m(\omega^n)=\delta_m^n$, defined by,
\bea
\begin{array}{rclcrcl}
\omega^0&=&A(\rho)\;dx^0
\;\;\;\;&,&\;\;\;\; e_0 &=& A(\rho)^{-1}\;\partial_0\cr
\omega^{I}&=&\tilde A(\rho)\;dx^I
\;\;\;\;&,&\;\;\;\; e_I &=& \tilde A (\rho)^{-1}\;\partial_I\qquad,\qquad I=1,\dots,p\cr
\omega^{p+1}&=&C(\rho)\;d\rho
\;\;\;\;&,&\;\;\;\; e_{p+1}&=& C(\rho)^{-1}\;\partial_{\rho}\cr
\omega^{p+2}&=&\tilde C(\rho)\; dz
\;\;\;\;&,&\;\;\;\; e_{p+2} &=& \tilde C(\rho)^{-1}\;\partial_z
\end{array}
\eea
\subsection{Connections.}

As usual, the connections $\omega^m{}_n$ are completely determined by
\begin{itemize}
\item  No torsion condition: \hspace{2cm}$d\omega^m+{\omega^m}_n\wedge \omega^n=0$
\item  Metricity:\hspace{4cm}$\omega_{mn}\equiv\eta_{mp}\,\omega^p{}_n =-\omega_{nm}.$
\end{itemize}
From the direct computation of the coefficients ${\;\alpha^m}_{nl}$,
\begin{equation}
d\omega^m \equiv {1\over2}{\alpha^m}_{n l}\;\omega^n \wedge
\omega^l\;\;\;,\;\;\; {\alpha^m}_{nl} = - {\alpha^m}_{ln}
\end{equation}
we get the connections in the form,
\be
\omega_{mn}\equiv\omega_{m n l}\;\omega^l\qquad,\qquad
\omega_{m n l}={1\over2}(\alpha_{mn l}-\alpha_{l m n}+\alpha_{n l m})
\ee
We resume the non zero connections associated with the metric in (\ref{ansatzapen}),
\begin{eqnarray}
\omega_{0 p+1}&=&{\partial_{\rho}\ln A\over C}\;\omega_0= e_{p+1}(\ln A)\;\omega_0 \cr
\omega_{I p+1}&=&{\partial_{\rho}\ln \tilde A\over C}\;\omega_I
=e_{p+1}(\ln\tilde A)\;\omega_I\cr
\omega_{p+1,p+2}&=&-{\partial_{\rho}\ln\tilde C\over C}\;\omega_{p+2}=
-e_{p+1}(\ln \tilde C)\;\omega_{p+2}\label{conexiones}
\end{eqnarray}

\subsection{Covariant derivatives.}

We collect here covariant derivatives of interest related to the connections presented before.
By definition,
\begin{center}
$D_m(A_n)\equiv e_m(A_n)-\omega_{s n m}\; A_s$
\end{center}
Because in our equations we have scalar functions of $\rho$, it is enough to know a few of them.
Let $\phi(\rho)$ be an arbitrary scalar function; then we have the following
non zero derivatives (contraction of indices is assumed),
\begin{eqnarray}
D_{p+1}(\phi) &=& e_{p+1}(\phi) = \frac{1}{C}\,\partial_{\rho}\phi\cr
D_0{}^2(\phi)&=&-\frac{1}{C^2}\;\partial_{\rho}\ln A\;\partial_{\rho}\phi\cr
D_I D_J (\phi)&=&\delta_{IJ}\;e_{p+1}(\ln\tilde A)\;e_{p+1}(\phi)\cr
D_{p+1}{}^2(\phi)&=&\partial_{\rho}\left({\partial_{\rho}\phi\over C^2}\right)
+ \frac{1}{C^2}\;\partial_{\rho}\ln C\;\partial_{\rho}\phi\cr
D_{p+2}{}^2(\phi)&=& \frac{1}{C^2}\;\partial_{\rho}\ln \tilde C \;\partial_{\rho}\phi\cr
D^2(\phi) &=& \frac{1}{F_1}\;\partial_{\rho}\left(\frac{F_1}{C^2}\;\partial_{\rho}\phi\right)
=\partial_{\rho}\left(\frac{\partial_{\rho}\phi}{C^2}\right)+
\frac{1}{C^2}\;\partial_{\rho}\ln F_1\;\partial_{\rho}\phi\cr
e^{b\xi}D(e^{-b\xi}\; \psi\; D(\phi)) &=&\frac{1}{F_1\, e^{-b\xi}}\;\partial_{\rho}\left(\frac{F_1\,e^{-b\xi}}{C^2}\;
\psi\;\partial_{\rho}\phi\right)\cr
& & \label{derivadascov}
\end{eqnarray}
where $F_1 \equiv A\,\tilde A^p\, C\,\tilde C\;$,
$\;\epsilon_G =  dx^0\wedge dx^1\wedge\dots\wedge dx^p\wedge d\rho\wedge
d\theta \;F_1\,$.

\subsection{Curvature tensor.}

By definition,
\be
\Re_{mn} \equiv d\omega_{mn} + \omega_{mp}\wedge\omega^p{}_n =
\frac{1}{2}\,\Re_{mnpq}\;\omega^p\wedge\omega^q
\ee
The computation yields,
\bea
\Re_{0I}&=& D_0{}^2(\ln \tilde A)\;\omega_0\wedge\omega_I\cr
\Re_{0p+1}&=&-{1\over{A}}{D_{p+1}}^2(A)\;\omega_0\wedge\omega_{p+1}\cr
\Re_{Ip+1}&=&-{1\over{\tilde{A}}}D_{p+1}{}^2(\tilde{A})\;\omega_I\wedge\omega_{p+1}\cr
\Re_{IJ}&=&- D_{p+1}(\ln\tilde A)^2\;\omega_I\wedge\omega_J\cr
\Re_{0p+2}&=&-\frac{1}{A}D_{p+2}{}^2(A)\;\omega_0\wedge\omega_{p+2}\cr
\Re_{Ip+2}&=&-\frac{1}{\tilde A}D_{p+2}{}^2(\tilde{A})\;\omega_I\wedge\omega_{p+2}\cr
\Re_{p+1p+2}&=&-{1\over{\tilde{C}}}D_{p+1}{}^2(\tilde C)\;\omega_{p+1}\wedge\omega_{p+2}
\label{curvatura}
\eea

\subsection{Ricci tensor and Ricci scalar.}

From (\ref{curvatura}), we get the following components of the Ricci tensor
$R_{mn} \equiv \Re^p{}_{mpn}$,
%\vfill\eject
\bea
R_{00}&=& D^2(\ln A)\cr
R_{IJ}&=&- D^2(\ln \tilde A)\;\delta_{IJ}\cr
R_{p+1p+1}&=&-D^2(\ln C)-{1\over A}\partial_{\rho}\left({\partial_{\rho}A\over
C^2}\right)-{p\over \tilde A}\partial_{\rho}
\left({\partial_{\rho}\tilde A\over C^2}\right)+
{1\over C}\partial_{\rho}\left({\partial_{\tilde
\rho}C\over C^2}\right)- {1\over \tilde C}\partial_{\rho}
\left({\partial_{\rho}\tilde C\over C^2}\right)\cr
&=&-D^2(\ln C) - \frac{D_{p+1}{}^2(A)}{A} - p\,\frac{D_{p+1}{}^2(\tilde A)}{\tilde A}
+\frac{D_{p+1}{}^2(C)}{C} -\frac{D_{p+1}{}^2(\tilde C)}{\tilde C}\cr
&+&\frac{1}{C^2}\;\partial_{\rho}\ln C\;\partial_{\rho}\ln \frac{F_1}{C^2}\cr
R_{p+2p+2}&=&- D^2(\ln \tilde C)
\label{ricci}
\eea
\bigskip
The curvature scalar results,
\bea
R&=&
-{2\over C}\partial_{\rho}^2\left({1\over C}\right)
-D(\ln A)^2-p\,D(\ln \tilde A)^2-D(\ln C)^2-D(\ln \tilde C)^2\cr
&-&2\,D^2(\ln F_1)+D(\ln F_1)^2
\eea
\label{ricciscalar}

\subsection{Strength field tensor.}

With the ans\"atz in (\ref{fieldansatz}) for the $p+1$ form we get,
\begin{eqnarray}
F_{p+2} &=& dx^0\wedge\dots\wedge dx^p\; \partial_\rho E(\rho)\cr
\left(F_{p+2}\right)^2_{\mu\nu}&=&-{(\partial_{\rho}E)^2\over(A\,\tilde{A}^p\,C)^2}\;
\eta_{\mu \nu}\cr
\left(F_{p+2}\right)^2_{p+1p+1}&=&-
{(\partial_{\rho}E)^2\over(A\,\tilde{A}^p\,C)^2}\;
\end{eqnarray}

The gauge contribution to the strength field tensor results,

\begin{eqnarray}
T^A_{\mu\nu}&=&-\frac{2-b_p}{8}\;\frac{e^{(2+ b_p)\Phi}}{\left(A\,{\tilde A}^p\,C\right)^2}\;
\; (\partial_{\rho}E)^2\;\eta_{\mu\nu}\cr
T^A_{p+1p+1}&=&-\frac{2-b_p}{8}\;\frac{e^{(2+ b_p)\Phi}}{\left(A\,{\tilde A}^p\,C\right)^2}\;
\; (\partial_{\rho}E)^2\cr
T^A_{p+2p+2}&=&\frac{2+b_p}{8}\;\frac{e^{(2+ b_p)\Phi}}{\left(A\,{\tilde A}^p\,C\right)^2}\;
\; (\partial_{\rho}E)^2\label{fieldstrength}
\end{eqnarray}

\subsection{The equations of motion.}

With the help of the precedent results for the Ricci tensor, strength field tensor, etc.,
the equations of motion (\ref{ecformal}) for the ans\"atz (\ref{fieldansatz}) can be recast in the following way,
\bigskip

%%%% A-equation
\noindent\underline{$A$-equation}
\be
e^{2\Phi}D(e^{-2\Phi}D(\ln A))={2-b_p\over 8}\;{e^{(b_p+2)\Phi}\over (A{\tilde A}^p\,C)^2}\;
(\partial_\rho E)^2\label{1}
\ee
%%%%% \tilde A-equation
\noindent\underline{$\tilde A$-equation}
\be
e^{2\Phi}D(e^{-2\Phi}D(\ln \tilde A))=
{2-b_p\over 8}\;{e^{(b_p+2)\Phi}\over(A{\tilde A}^p\,C)^2}\;(\partial_\rho E)^2\label{2}
\ee
%%%% C-equation
\noindent\underline{$C$-equation}
\bea
&-&e^{2\Phi}D(e^{-2\Phi}D(\ln C))={D_{p+1}{}^2(A)\over A} + p\,{D_{p+1}{}^2(\tilde A)\over
\tilde A}- {D_{p+1}{}^2(C)\over C} + {D_{p+1}{}^2(\tilde C)\over \tilde C}\cr
&-&2\,D_{p+1}{}^2(\Phi)
-\frac{1}{C^2}\; \partial_\rho\ln C\;\partial_\rho\ln\left({F_1\over C^2}e^{-2\Phi}\right)
- \frac{2-b_p}{8}\;{e^{(b_p+2)\Phi}\over (A{\tilde A}^p\,C)^2}\;(\partial_\rho E)^2\label{3}
\eea
%%%%%%%% \tilde C-equation
\noindent\underline{$\tilde C$-equation}
\be
-e^{2\Phi}D(e^{-2\Phi}D(\ln \tilde C))={2+b_p\over 8}\;
{e^{(b_p+2)\Phi}\over (A{\tilde A}^p\,C)^2}\;(\partial_\rho E)^2\label{4}
\ee
%%%%%%%% Phi-equation
\noindent\underline{$\Phi$-equation}
\be
e^{2\Phi}D(e^{-2\Phi}D(\ln e^{-2\Phi}))=\Lambda^2 - (p+1)\,{2-b_p\over 8}\;
{e^{(b_p+2)\Phi}\over (A\tilde A^p\,C)^2}\;(\partial_\rho E)^2\label{5}
\ee
%%%% E-equation
\noindent\underline{E-equation}
\be
-D\left(\frac{e^{b_p\Phi}}{(A{\tilde A}^p)^2}\;D(E)\right)=
(-)^p \,Q_p\;\frac{\delta^2_{\perp}}{F_1}\label{6}
\ee
In the last one, the $E$-equation, we used the relation,
\be
*F_{p+2} = (-)^p\;\frac{\tilde C^2}{F_1}\;\partial_{\rho}E\;dz
\ee
and assumed a source of the form $\,Q_p\,J_{p+1}$,
\be
J_{p+1}^{01\dots p} = \frac{\delta^2_\perp}{\sqrt{-\det G}}=
\frac{\delta^2_{G^\perp}}{A\,\tilde A^p}\;\;\longrightarrow\;\;
J_{p+1} = -\delta^2_{G^\perp}\;\omega^0\wedge\dots\wedge\omega^p
\ee
where $G^\perp$ stands for the metric of the space where the flat $p$ brane is localized.
%\vfill\eject
%%%%%%%%%%%%%%%%%%%%%%%%%%%%%%%%%%%%%%%%%%%%%%%%%%%%%%%%%%%%%%%%%%%%%%%%%%%%%%%%%
%%%%%%%%%%%%%%%%%%%%%%%%%%% BIBLIOGRAPHY %%%%%%%%%%%%%%%%%%%%%%%%%%%%%%%%%%%%%%%%%%%%%%%
%%%%%%%%%%%%%%%%%%%%%%%%%%%%%%%%%%%%%%%%%%%%%%%%%%%%%%%%%%%%%%%%%%%%%%%%%%%%%%%%%

%%%%%%%%%%%%%%%%%%%%%%%%%%%%%%%%%%%%%%%%%%%%%%%%%%%%%%%%%%%%%%%%%%%%%%%%%%%%%%%%%
%%%%%%%%%%%%%%%%%%%%%%%%%%% END BIBLIOGRAPHY %%%%%%%%%%%%%%%%%%%%%%%%%%%%%%%%%%%%
%%%%%%%%%%%%%%%%%%%%%%%%%%%%%%%%%%%%%%%%%%%%%%%%%%%%%%%%%%%%%%%%%%%%%%%%%%%%%%%%%

\end{document}